\documentclass[aps,prd,groupedaddress,showpacs,showkeys]{revtex4}
\usepackage{amssymb}
\usepackage{amsmath}
\usepackage{amsfonts}
\usepackage{epsfig}
\usepackage{amssymb}
\usepackage{amsmath}
\usepackage{amsfonts}
\usepackage{epsfig}

\newcommand{\seq}{\begin{subequations}}
\newcommand{\sen}{\end{subequations}}
\newcommand{\eq}{\begin{eqnarray}}
\newcommand{\en}{\end{eqnarray}}

\newcommand{\la}{\langle} 
\newcommand{\ra}{\rangle} 
\newcommand{\beqn}{\begin{eqnarray}}
\newcommand{\eeqn}{\end{eqnarray}}
\newcommand{\beq}{\begin{equation}}
\newcommand{\eeq}{\end{equation}}
\newcommand{\barr}{\begin{array}}
\newcommand{\earr}{\end{array}}

\begin{document}

\title{Semileptonic decays of double heavy baryons \\
in a relativistic constituent three--quark model} 
\noindent
\author{Amand Faessler$^{1}$,
        Thomas Gutsche$^{1}$,
        Mikhail A. Ivanov$^{2}$,  
        J\"urgen G. K\"{o}rner$^{3}$, 
        Valery E. Lyubovitskij$^1$\footnote{On leave of absence
          from Department of Physics, Tomsk State University,
          634050 Tomsk, Russia}
\vspace*{1.2\baselineskip}}

\affiliation{$^1$ Institut f\"ur Theoretische Physik,
Universit\"at T\"ubingen,\\ 
Kepler Center for Astro and Particle Physics, \\
Auf der Morgenstelle 14, D--72076 T\"ubingen, Germany
\vspace*{1.2\baselineskip} \\
$^2$ Bogoliubov Laboratory of Theoretical Physics,
Joint Institute for Nuclear Research,~141980~Dubna,~Russia
\vspace*{1.2\baselineskip} \\
$^3$ Institut f\"{u}r Physik, Johannes Gutenberg-Universit\"{a}t,
D--55099 Mainz, Germany \\}

\date{\today}

\begin{abstract}

We study the semileptonic decays of double heavy baryons using 
a manifestly Lorentz covariant constituent three--quark model. 
We present complete results on transition form factors between 
double--heavy baryons for finite values of the heavy quark/baryon 
masses and in the heavy quark symmetry limit which is valid at 
and close to zero recoil. Decay rates are calculated and compared 
to each other in the full theory, keeping masses finite,  
and also in the heavy quark limit. 

\end{abstract}



\maketitle

\newpage

\section{Introduction}

The semileptonic decays of double--heavy baryons provide yet another 
opportunity to measure the Cabibbo--Kobayashi--Maskawa (CKM) matrix element 
$V_{cb}$. This is particularly so since the transition matrix elements between
double--heavy baryons obey spin symmetry relations in the heavy quark limit 
in addition to a model independent zero recoil normalization of the relevant 
transition matrix elements. In this paper we study current--induced
transitions between double--heavy baryons in a fully relativistic
constituent three--quark model. In the heavy quark limit we recover the
spin symmetry relations among the form factors valid at zero recoil and 
close to zero recoil including their zero recoil
normalization. Since the model is formulated in terms of finite values of the 
quark and baryon masses we are able to calculate the corrections to the spin 
symmetry relations and the zero recoil normalization valid in the heavy quark 
limit. 

Current--induced double--heavy baryon transitions
have been analyzed in a number of model approaches. These include  
effective field theories based on heavy quark spin 
symmetry~\cite{White:1991hz,SanchisLozano:1994vh,%
Flynn:2007qt,Hernandez:2007qv}, 
three--quark models ~\cite{Faessler:2001mr,Albertus:2006ya,Roberts:2008wq}, 
quark--diquark models~\cite{Guo:1998yj,Ebert:2004ck}, and 
nonrelativistic QCD sum rules~\cite{Onishchenko:2000wf,Kiselev:2001fw}. 
The progress achieved up to now can be summarized as follows. In the 
heavy quark limit (HQL) or in the limit of infinitely heavy quarks 
($m_{c,b} \to \infty$) double--heavy baryons can be viewed as heavy--light 
meson--like states containing a heavy diquark in the $\bar 3$ color 
state and a light quark in the $3$ color state~\cite{White:1991hz}. In the 
HQL the spins of the light quark and heavy diquark system decouple. This 
gives rise to relations between different transition form factors
involving double--heavy baryons in the quark--diquark 
picture~\cite{White:1991hz} as well as in the
three--quark picture~\cite{Flynn:2007qt}. In particular, working 
in the near zero recoil limit one can express all weak transition form 
factors between double heavy baryons through a single universal function 
$\eta(\omega)$ which depends on the kinematical parameter $\omega=v \cdot v'$ 
where $v$ and $v'$ stand for the four--velocities of the initial and final
double--heavy baryon, respectively. 

In Ref.~\cite{Faessler:2001mr} we have analyzed double heavy baryons 
for specific decay modes. We have restricted ourselves to spin $1/2$ to 
spin $1/2$ transitions using the same relativistic constituent three--quark 
model (RTQM)~\cite{Faessler:2001mr,Ivanov:1996pz} as is being used in the
present paper except that we now no longer have to rely on the impulse 
approximation. 
Differing from the approach of the present
paper, in~\cite{Faessler:2001mr} we have treated the double--heavy baryons
as bound states of a heavy $b$-quark and a heavy--light (cq) diquark. In this
paper we take double--heavy baryons to be bound states of a light quark and a 
double--heavy (bc) diquark. In particular this means that the interpolating
three--quark currents used in this paper have a different spin--flavor 
structure than the corresponding current in~\cite{Faessler:2001mr}. 
In the full theory this will lead to different predictions for the rates. 
We would like to emphasize, though, that, in the nonrelativistic limit, 
both currents are consistent with one another.  

The RTQM can be viewed as an effective 
quantum field theory approach based on an interaction Lagrangian of hadrons  
interacting with their constituent quarks. From such an approach one can
derive universal and reliable predictions 
for exclusive processes involving both mesons composed of a quark and 
antiquark and baryons composed of three quarks. 
The coupling strength of a hadron $H$  to its constituent quarks
is determined by the compositeness condition 
$Z_H=0$~\cite{Weinberg:1962hj,Efimov:1993ei} where 
$Z_H$ is the wave function renormalization constant of the hadron. The 
quantity $Z_H^{1/2}$ is the matrix element between a physical particle state 
and the corresponding bare state. The compositeness condition $Z_H=0$ enables 
one to represent a bound state by introducing a hadronic field interacting 
with its constituents so that the renormalization factor is equal to zero. 
This does not mean that we can solve the QCD bound state equations but we are 
able to show that the condition $Z_H=0$ provides an effective and 
self--consistent way to describe the coupling of a hadron to its 
constituents. One  starts with an effective interaction Lagrangian written 
down in terms of quark and hadron variables. Then, by using Feynman rules, 
the $S$--matrix elements describing hadron-hadron interactions are given in 
terms of a set of quark level Feynman diagrams. In particular, the 
compositeness condition enables one to avoid the problem of double counting of
quark and hadronic degrees of 
freedom. The approach is self--consistent and all calculations of physical 
observables are straightforward. There is  a small set of  model parameters: 
the values of the constituent quark masses and the scale parameters that 
define the size of the distribution of the constituent quarks inside a given 
hadron. 

The main objective of the present paper is to present 
a comprehensive analysis of all possible current--induced spin transitions  
between double--heavy baryons containing both types of light quarks -- 
nonstrange $q=u,d$ and strange $s$. This involves the flavor transitions
$bc \to cc$ and $bb \to bc$ where the transition
$bc \to cc$ is treated as the generic process in the main text while 
the results for the transition $bb \to bc$ are mainly relegated to
tables. The paper is structured as follows. 
First, in Sec.II we present interpolating three--quark currents with 
the appropriate quantum numbers 
of the double--heavy baryons. We then write down the corresponding Lagrangians 
defining the couplings of these currents to double--heavy baryons. 
Second, we briefly discuss the calculational techniques of how to calculate 
transition matrix 
elements generated by the Lagrangian functions. In Sec.III we consider the 
heavy quark limit of our transition matrix elements and recover the known
heavy quark symmetry relations for the transition matrix elements 
between double heavy baryons as well as the appropriate zero recoil 
normalization of the form factors. In particular,
we compare the results of the full finite mass calculation with the results
derived in the HQL. Fourth, in Sec.IV we present our numerical results which 
are compared to predictions of other theoretical approaches. In particular, 
we compare the results of the full finite mass calculation with 
the results derived in the HQL.
Finally, in Sec.~IV we present a short summary of our results.

\section{Semileptonic decays of double--heavy baryons}

\subsection{Lagrangian}

For the evaluation of the semileptonic decays we will consistently 
employ the relativistic constituent three--quark model. 
In the following we present details of the model which is
based on an interaction Lagrangian describing
the coupling between baryons and their constituent quarks.

The coupling of a baryon $B(q_1q_2q_3)$ to its constituent 
quarks $q_1$, $q_2 $ and $q_3$ is described by the Lagrangian 
\eq\label{Lagr_str}
{\cal L}_{\rm int}^{\rm str}(x) = g_B \bar B(x) \, 
\int\!\! dx_1 \!\! \int\!\! dx_2 \!\! \int\!\! dx_3 \, 
F_B(x,x_1,x_2,x_3) \, J_B(x_1,x_2,x_3) \, + \, {\rm h.c.}  
\en 
where $J_{B}(x_1,x_2,x_3)$ is the interpolating three--quark current 
with the quantum numbers of the relevant baryon $B$. One has
\eq 
J_{B}(x_1,x_2,x_3) \, = \, \varepsilon^{a_1a_2a_3} \, 
\Gamma_1 \, q^{a_1}_1(x_1) \, q^{a_2}_2(x_2) 
C \, \Gamma_2 \, q^{a_3}_3(x_3) \, ,
\label{lag}
\en 
where the $\Gamma_{1,2}$ are sets of Dirac matrices, $C$ is the charge 
conjugation
matrix $C=\gamma^{0}\gamma^{2}$ and the $a_i$ (i=1,2,3) are color indices.
$F_B(x,x_1,x_2,x_3)$ is a nonlocal scalar vertex function which characterizes
the finite size of the baryon.

\subsection{Vertex function}

The vertex function $F_B$ is related to the scalar part of the 
Bethe--Salpeter amplitude and characterizes the finite size 
of the baryon. To satisfy translational invariance the function 
$F_B$ has to fulfill the identity 
\eq\label{trans_inv}
F_B(x+a,x_1+a,x_2+a,x_3+a) \, = \, 
F_B(x,x_1,x_2,x_3) 
\en
for any given four--vector $a\,$. 
In the following we use a specific form for the vertex function 
\eq\label{vertex}
F_B(x,x_1,x_2,x_3) \, = \, N_B \, 
\delta^{(4)}(x - \sum\limits_{i=1}^3 w_i x_i) \;  
\Phi_B\biggl(\sum_{i<j}( x_i - x_j )^2 \biggr) 
\en 
where $\Phi_B$ is a nonlocal correlation function involving the three 
constituent quarks with masses $m_1$, $m_2$, $m_3$; $N_B = 9$ is a 
normalization factor. The variable $w_i$ is defined by 
$w_i=m_i/(m_1+m_2+m_3)$. The vertex function (\ref{vertex}) 
satisfies the translational identity (\ref{trans_inv}). 

The Fourier transform of the correlation function
$\Phi_B\biggl(\sum\limits_{i<j} ( x_i - x_j )^2 \biggr)$ can be calculated
by using Jacobi coordinates. One has
\begin{eqnarray}
\label{fourier}
{\widetilde{\Phi}}_{B}(p_1,p_2,p_3) &=& N_B \int dx e^{-ipx} 
\prod\limits_{i=1}^3\int\! dx_i e^{ip_ix_i}\, 
\delta^{(4)}(x - \sum\limits_{i=1}^3 w_i x_i) \, 
\Phi_B\biggl( \sum_{i<j}( x_i - x_j )^2 \biggr)  
\nonumber\\
&=& (2\pi)^4\,\delta^{(4)}\Big(p  - \sum\limits_{i=1}^3 p_i\Big)\,
{\overline{\Phi}}_{B}(-l_1^2-l_2^2)\,,
\end{eqnarray}
where the  Jacobi coordinates are defined by  
\begin{eqnarray}
x_1&=&x \, + \, \tfrac{1}{\sqrt{2}}\, \xi_1 w_3 
        \, - \, \tfrac{1}{\sqrt{6}}\, \xi_2 (2 w_2 + w_3)\,,
\nonumber\\
x_2&=&x \, + \, \tfrac{1}{\sqrt{2}}\, \xi_1  w_3      
        \, + \, \tfrac{1}{\sqrt{6}}\, \xi_2 (2 w_1 + w_3)\,, 
\nonumber\\ 
x_3&=&x \, - \, \tfrac{1}{\sqrt{2}}\, \xi_1  (w_1  + w_2)     
        \, + \, \tfrac{1}{\sqrt{6}}\,  \xi_2 (w_1  - w_2)\,.
\end{eqnarray}
The corresponding Jacobi momenta read
\begin{eqnarray}
p &=& p_1+p_2+p_3\,,
\nonumber\\
l_1 &=& \tfrac{1}{\sqrt{2}}\,w_3 (p_1 + p_2) 
      - \tfrac{1}{\sqrt{2}}\,(w_1+w_2) p_3\,, 
\nonumber\\
l_2 &=&  - \tfrac{1}{\sqrt{6}}\,(2 w_2 + w_3) p_1
         + \tfrac{1}{\sqrt{6}}\,(2 w_1 + w_3) p_2
         + \tfrac{1}{\sqrt{6}}\,(w_1 - w_2) p_3\,, 
\end{eqnarray} 
where, according to Eq.(\ref{vertex}), $\sum\limits_{i=1}^3w_ix_i=x$. Since 
the function
$\Phi_B\biggl(\sum\limits_{i<j}( x_i - x_j )^2 \biggr)$
is invariant under translations, its Fourier transform only 
depends on two four--momenta. 
The function ${\overline{\Phi}}_{B}(-l_1^2-l_2^2)$ in Eq.~(\ref{fourier}) 
will be modelled in our approach. 
The minus sign in the argument is chosen to emphasize
that we are working in Minkowski space.
A simple choice is the Gaussian form 
\eq\label{corr_Fun}
{\overline{\Phi}}_{B}(-l_1^2-l_2^2) = \exp(18 (l_1^2+l_2^2)/\Lambda_B^2) 
\en 
where the parameter $\Lambda_B$ characterizes the size of the double heavy 
baryon. Since $l_1^2$ and $l_2^2$ turn into $-l_1^2$ and $-l_2^2$ in 
Euclidean space the form~(\ref{corr_Fun}) has the appropriate fall--off
behavior in the Euclidean region. 

\subsection{Three--quark currents}

Double--heavy baryons  are classified by the set of quantum numbers 
$(J^P, S_d)$, where $J^P$ is the spin--parity of the baryon state 
and $S_d$ is the spin of the heavy diquark. There are two types of heavy 
diquarks -- those with $S_d = 0$ 
(antisymmetric spin  configuration $[Q_1Q_2]$) and those with $S_d = 1$ 
(symmetric spin 
configuration $\{Q_1Q_2\}$). Accordingly there are two $J^{P}=1/2\,^+$ 
double--heavy baryon states. We follow the standard convention and attach 
a prime to the $S_d = 0$ states whereas the 
$S_d = 1$ states are unprimed. Note that the 
$J^P = 3/2\,^+$ states are in the symmetric heavy quark spin 
configuration. In Table 1 we list the  
quantum numbers of the double--heavy baryons including their mass spectrum as 
calculated in~\cite{Ebert:2004ck}.

We pause for a moment to discuss some of the features of the mass spectrum
obtained in~\cite{Ebert:2004ck} which are relevant for our calculation.
One notes that there is a mass inversion in the $(1/2^+)$ mixed flavor states
$(\Xi_{bc},\Xi'_{bc})$ and $(\Omega_{bc},\Omega'_{bc})$ in that
$M(\Xi'_{bc})>M(\Xi_{bc})$ and $M(\Omega'_{bc})>M(\Omega_{bc})$ contrary
to naive expectation even though the heavy triplet diquark state has a higher 
mass than the heavy singlet diquark state in the
model of~\cite{Ebert:2004ck}, i.e. $m(bc;S_{d}=1) >  m(bc;S_{d}=0)$. The
inverted mass hierarchy is at the origin of the prime notation mentioned
above. We mention that the
inverted mass hierarchy is a feature of all models that have attempted to
calculate the mass spectrum of double--heavy baryons~\cite{Albertus:2006ya,
Ebert:2004ck,Kiselev:2001fw,Roberts:2007ni,Zhang:2008pm,Bernotas:2008fv}. 
In particular, the
inverted mass hierarchy implies that one can only expect substantial 
flavor--changing branching ratios for the two lowest lying states $\Xi_{bc}$ 
and $\Omega_{bc}$ whereas the rates of the higher lying states $\Xi_{bc}'$ and
$\Xi_{bc}^{*}$, and $\Omega_{bc}'$ and $\Omega_{bc}^{*}$ will be dominated
by flavor--preserving one-photon transitions to the lowest-lying states
$\Xi_{bc}$ and $\Omega_{bc}$. It will be interesting to analyze the strength 
of one--photon
transitions between the $S_d = 0$ and $S_d = 1$ double--heavy baryon states
which are forbidden in the HQL since, in the HQL, the photon couples to
the light quark only. For finite heavy quark masses one-photon transitions
between the $S_d = 0$ and $S_d = 1$ double--heavy baryon states will occur at 
a somewhat
reduced rate which, however, very likely will still exceed the 
flavor--changing weak decay rates of these states.   
 
We construct the interpolating currents of the double heavy baryon 
$B_{qQ_1Q_2}$ 
in the form of a light quark $q^{a_1}$ coupled to a heavy diquark 
$d^{a_1}_{Q_1Q_2}$, {\it viz}. 
\eq 
J_{qQ_1Q_2} &=& \Gamma_{Q_1Q_2} q^{a_1}d^{(Q_1Q_2)}_{a_1}, \qquad
d^{(Q_1Q_2)}_{a_1}=\varepsilon^{a_1a_2a_3} \ \Big(Q_{1}^{a_2} 
C \Gamma^{(d)}_{Q_1Q_2} Q_2^{a_3}\Big) \,. 
\en 
We shall only consider currents without derivatives. With this restriction
one can construct three interpolating currents for the $(\frac{1}{2}^+, 0)$ 
states -- the pseudoscalar $J^P$, 
scalar $J^S$ and axial $J^A$ currents
\seq 
\eq 
J_{qQ_1Q_2}^P &=& \varepsilon^{a_1a_2a_3} \
q^{a_1} \Big(Q_{1}^{a_2} 
C \gamma_5 Q_2^{a_3}\Big)\,, \\ 
J_{qQ_1Q_2}^S &=& \varepsilon^{a_1a_2a_3} \ 
\gamma^5 q^{a_1} \Big(Q_{1}^{a_2} 
C Q_2^{a_3}\Big)\,, \\
J_{qQ_1Q_2}^A &=& \varepsilon^{a_1a_2a_3} \ 
\gamma^\mu q^{a_1} \Big(Q_{1}^{a_2} 
C \gamma_5 \gamma_\mu Q_2^{a_3}\Big)\,. 
\en 
\sen 
For the $(\frac{1}{2}^+, 1)$ states one has 
a vector $J^V$ and a tensor $J^T$ current  
\seq 
\eq 
J_{qQ_1Q_2}^V &=& \varepsilon^{a_1a_2a_3} \ 
\gamma^\alpha\gamma^5  q^{a_1} \Big(Q_{1}^{a_2} 
C \gamma_\alpha Q_2^{a_3}\Big)\,,  \\
J_{qQ_1Q_2}^T &=& \frac{1}{2} \varepsilon^{a_1a_2a_3} \  
\sigma^{\mu\nu} \gamma^5  q^{a_1} \Big(Q_{1}^{a_2} 
C \sigma_{\mu\nu} Q_2^{a_3}\Big)\,. 
\en  
\sen 
Finally, for the $(\frac{3}{2}^+, 1)$ states one has 
the vector and tensor currents $J^V_\mu$ and $J^T_\mu$ 
\seq 
\eq 
J_{qQ_1Q_2, \ \mu}^V &=& \varepsilon^{a_1a_2a_3} \ 
q^{a_1} \Big(Q_{1}^{a_2} 
C \gamma_\mu Q_2^{a_3}\Big)\,, \\
J_{qQ_1Q_2,  \ \mu}^T &=& - i \varepsilon^{a_1a_2a_3} \ 
\gamma^\nu q^{a_1} \Big(Q_{1}^{a_2} 
C \sigma_{\mu\nu} Q_2^{a_3}\Big)\,.  
\en  
\sen 
Note that any double--heavy baryon current in the form of a heavy 
quark coupling to a heavy--light diquark can be transformed to a linear
combination of the above form of currents using a Fierz transformation. 

In the heavy quark limit the scalar current $J_{Q_1Q_2q}^S$ 
vanishes, while the other currents become degenerate in the following way: 
\seq 
\eq 
J_{qQ_1Q_2}^P \ = \ J_{qQ_1Q_2}^A &=& \varepsilon^{a_1a_2a_3} \ 
\psi_{q}^{a_1} \ ( \psi_{Q_1}^{a_2} \, \sigma_2 \, \psi_{Q_2}^{a_3} ) \,, \\
J_{qQ_1Q_2}^V \ = \ J_{qQ_1Q_2}^T &=& \varepsilon^{a_1a_2a_3} \ \vec{\sigma} 
\, \psi_{q}^{a_1} \ ( \phi_{Q_1}^{a_2} \, \sigma_2 \vec{\sigma} 
\, \psi_{Q_2}^{a_3} ) \,, \\ 
\vec{J}_{qQ_1Q_2}^{\, V} \ = \ \vec{J}_{qQ_1Q_2}^{\, T} 
&=& \varepsilon^{a_1a_2a_3} \ \psi_{q}^{a_1} \ 
( \psi_{Q_1}^{a_2} \, \sigma_2 \vec{\sigma} \, \psi_{Q_2}^{a_3} ) \,, 
\en 
\sen 
where $\psi_{q,Q_1,Q_2}$ are the upper components of the Dirac quark spinors  
and the $\sigma_i$ are Pauli spin matrices. Excluding the scalar
current $J_{Q_1Q_2q}^S$, which vanishes in the HQL, we remain with two
currents for each of the double--heavy baryon states which, as shown above, 
become degenerate in the HQL. It is therefore reasonable to take only one of 
the interpolating currents each. Our choice is to take 
the simplest current from each pair -- 
the pseudoscalar current for the $(\frac{1}{2}^+, 0)$ states and  
the vector currents for the $(\frac{1}{2}^+, 1)$ and $(\frac{3}{2}^+, 1)$ 
states. Note that the HQL coincides with the nonrelativistic limit. 
In the nonrelativistic limit our double heavy baryon (DHB) currents have 
a one--to--one correspondence to the naive quark model baryon spin-flavor 
functions which are displayed in Table 2. Further details on the naive 
quark model and how to evaluate the semileptonic current-induced transition 
amplitudes in this framework can be found in Appendix~A. 

We now shall give explicit expressions for the three--quark
currents which are needed for the calculation
of the $bcq \to ccq$ semileptonic transition amplitudes: 
\seq 
\eq 
J_{bcq} &=& \Gamma^{(q)}_{bc}q^{a_1}d_{(bc)}^{a_1}, \qquad
d_{(bc)}^{a_1}=\varepsilon^{a_1a_2a_3}\Big(b^{a_2} 
C \Gamma^{(d)}_{bc} c^{a_3}\Big)\,,\\
{\bar J}_{bcq} &=& {\bar d}_{(bc)}^ {a_1} {\bar q}^{a_1}
{\overline \Gamma}^{(q)}_{bc}, \qquad
{\bar d}_{(bc)}^{a_1}=\varepsilon^{a_1a_2a_3}
\Big({\bar c}^{a_2} {\overline \Gamma}^{(d)}_{bc}C {\bar b}^{a_3}\Big)
\en 
\sen 
where ${\bar J} = J^\dagger\gamma^0$ and 
$\overline\Gamma=\gamma^0\Gamma^\dagger\gamma^0$.
The corresponding $ccq$--currents are obtained by the obvious replacement
$b\to c$. Note that $C\Gamma^{(d)}_{cc}C = -\Gamma^{(d)\,T}_{cc}$.
Using a rather suggestive notation we specify the coupling content of the
DHB currents for the $\Xi_{Q_{1}Q_{2}}$ baryons using the form 
$\Longrightarrow \Gamma^{(q)} \otimes  \Gamma^{(d)} $. We thus consider the
currents 
\seq 
\eq 
\Xi_{cc} &\Longrightarrow& 
\gamma^{\alpha}\gamma^5 \otimes \gamma_{\alpha}\,, \qquad
\Xi^\ast_{cc} \Longrightarrow  I \otimes \gamma^{\nu}  \,,
\\
 \Xi_{bc} &\Longrightarrow& 
\gamma^{\alpha}\gamma^5 \otimes \gamma_{\alpha}\,, \qquad
\Xi^\ast_{bc} \Longrightarrow I \otimes \gamma^{\nu}\,, \qquad
\Xi^\prime_{bc} \Longrightarrow I \otimes \gamma^{5} .
\en 
\sen 
We use the same set of currents for the double--heavy 
$\Omega_{Q_{1}Q_{2}}$-type
baryons replacing the light $u,d$--quark by a $s$--quark.

\subsection{Normalization}

As described e.g. in~\cite{Faessler:2001mr} we need the derivative
of the mass operator for the double--heavy baryon $\Xi_{Q_1Q_2}$ in order
to evaluate the coupling constants $g_{B}=g_{\Xi_{Q_1Q_2}}$. 
The mass operator is given by a two--loop Feynman diagram and reads
\eq
\label{massop} 
\widetilde\Pi_{\, \Xi_{Q_1Q_2}}(p) &=& \frac{N_{q_1q_2}}{(16\pi^2)^2}
\int\!\frac{d^4k_1}{\pi^2 i}\int\!\frac{d^4k_2}{\pi^2 i}\,\,
\overline\Phi^2_B(- l_1^2 - l_2^2) \nonumber\\  
&\times&{\rm tr}\Big[\Gamma^{(d)}\tilde S_1(k_1+w_1 p)
{\bar \Gamma}^{(d)}\tilde S_2(k_2-w_2 p)\Big]
\ 
\Gamma^{(q)}\tilde S_3(k_2-k_1+w_3 p){\bar \Gamma}^{(q)},
\nonumber\\
&&\\
l_1 &=& \tfrac{1}{\sqrt{2}} \, (k_1 - k_2) \,, \qquad
l_2 \ = \ - \tfrac{1}{\sqrt{6}}  \, (k_1 + k_2) \,, \nonumber 
\en 
where $q_1=b$ or $c$, $q_2=c$ , $q_3=q$ and $N_{bc}=6$,  $N_{cc}=12$. 
The expression $\tilde S_i(k) = (m_{q_i} - \not\! k - i\epsilon)^{-1}$ 
denotes the free fermion propagator for the constituent quark with mass 
$m_{q_i}$. Integration momenta have been shifted in such a way so as to
remove the external momentum from the vertex function.
We have assigned outgoing momenta to the first and third outgoing quarks
and an in--going momentum for the second
quark.
Since the $p$--dependence of the mass operator resides entirely in the
propagators it is not difficult to calculate the derivative of the mass 
operator needed for the normalization condition 
\eq 
g^2_{\, \Xi_{Q_1Q_2}} \frac{d}{dp^\mu}
\widetilde\Pi_{\,\Xi_{q_1q_2}}(p) = \gamma^\mu\,.
\en 
The latter condition is known as a Ward identity which is 
equivalent to the compositeness condition
$Z_H=0$~\cite{Weinberg:1962hj,Efimov:1993ei}.   
One obtains
\eq 
\frac{d}{dp^\mu}\widetilde\Pi_{\, \Xi_{Q_1Q_2}}(p)  &=& 
\frac{N_{q_1q_2}}{(16\pi^2)^2}
\int\!\frac{d^4k_1}{\pi^2 i}\int\!\frac{d^4k_2}{\pi^2 i}\,\,
\overline\Phi^2_B(- l_1^2 - l_2^2) 
\nonumber\\
&\times&
\Big\{ w_1\,{\rm tr}\Big[
        \Gamma^{(d)}\tilde S_1(k_1+w_1 p)\gamma^\mu \tilde S_1(k_1+w_1 p)
        {\bar \Gamma}^{(d)}\tilde S_2(k_2-w_2 p)\Big]
\nonumber\\
&\times& \Gamma^{(q)}\tilde S_3(k_2-k_1+w_3 p){\bar \Gamma}^{(q)}
\nonumber\\
&-&\,w_2\,  {\rm tr}\Big[
       \Gamma^{(d)}\tilde S_1(k_1+w_1 p) {\bar \Gamma}^{(d)}
       \tilde S_2(k_2-w_2 p)\gamma^\mu \tilde S_2(k_2-w_2 p)\Big]
\label{norm}\\
&\times&   \Gamma^{(q)}\tilde S_3(k_2-k_1+w_3 p)){\bar \Gamma}^{(q)}
\nonumber\\
&+&\, w_3\,  {\rm tr}\Big[
       \Gamma^{(d)}\tilde S_1(k_1+w_1 p) {\bar \Gamma}^{(d)}
       \tilde S_2(k_2-w_2 p)\Big]
\nonumber\\
&\times&  
      \Gamma^{(q)} \tilde S_3(k_2-k_1+w_3 p)\gamma^\mu 
\tilde S_3(k_2-k_1+w_3 p)
      {\bar \Gamma}^{(q)} 
\Big\}\,, \nonumber  
\en 
where the double--heavy $\Xi_{Q_{1}Q_{2}}$--baryon is taken to be on its 
mass--shell. 

\subsection{Matrix elements of semileptonic decays}

The generic matrix element describing the semileptonic
transitions $\Xi_{bc} \to \Xi_{cc}$ reads 
\eq 
\la\Xi_{cc}(p')| \int d^4x e^{iqx} 
\bar c(x) O^\mu b(x)|\Xi_{bc}(p)\ra
=(2\pi)^4\,\delta^{(4)}(p-p'-q)\, \bar u_{\Xi_{cc}}(p^\prime) 
\Lambda^\mu(p,p') u_{\Xi_{bc}}(p)\,,  
\en 
where 
\eq 
\label{transition}
\Lambda^\mu(p,p') &=& 
12\,\frac{ g_{\,\Xi_{bc}}\,g_{\,\Xi_{cc}} }{(16\pi^2)^2}\,
\int\frac{d^4k_1}{\pi^2 i}\int\frac{d^4k_2}{\pi^2 i}
\overline\Phi_B(- l_1^2 - l_2^2) 
\overline\Phi_B(- l_1^{\prime\,2} - l_2^{\prime\,2}) 
\nonumber\\
&\times&{\rm tr}\Big[
\Gamma^{(d)}_{cc}\tilde S_4(k_1+\tfrac12\, p')\,O^\mu\,
\tilde S_1(k_1+p-\tfrac12\, p') \bar\Gamma^{(d)}_{bc}
\tilde S_2(k_2-\tfrac12\, p')\Big] \ 
\Gamma^{(q)}_{cc}\tilde S_3(k_2-k_1) \bar\Gamma^{(q)}_{bc} \,. 
\label{semlep}
\en 
Here $O^\mu=\gamma^\mu(1-\gamma^5)$ and
$w'_i=m_i/(m_4+m_2+m_3)$ with $i=4,2,3$. For the masses one has
$m_1=m_b$, $m_2=m_4=m_c$, $m_3=m_q$.
The Jacobi momenta $\omega_i$ and $\omega'_i$ are chosen as 
\eq 
l_1 &=& \tfrac{1}{\sqrt{2}}\,\Big[k_1 - k_2 + w_3 p\Big],
\nonumber\\
l_2 &=& - \tfrac{1}{\sqrt{6}}\,\Big[k_1 + k_2 + (w_2-w_1) p + q\Big],
\nonumber\\
&&\\
l'_1 &=& - \tfrac{1}{\sqrt{2}}\,\Big[k_1 - k_2 + w_3' p'\Big],
\nonumber\\
l'_2 &=& \tfrac{1}{\sqrt{6}}\,\Big[k_1 + k_2 + (w_2' - w_4') p'\Big]\,. 
\nonumber
\en 
Note that the expressions for the normalization and the vertex given in
Eq.~(\ref{norm}) and Eq.~(\ref{semlep}), resp., are exact in the sense that 
they are obtained directly from the Lagrangian Eq.~(\ref{lag})
for an arbitrary translationally invariant vertex function $F_B$ such as the
one defined in 
Eq.~(\ref{vertex}).
The two--loop integrals in Eqs.~(\ref{norm},\ref{semlep}) are invariant under
translations of the loop--variables $k_i\to k_i + b_i$ (i=1,2) where
$b_i$ are arbitrary momentum four--vectors. We have assigned the loop momenta 
such that the heavy quark limit can easily be taken. 
Calculational techniques for the two--loop quark integrals 
are given in some detail in our previous publication~\cite{Faessler:2001mr}.
One uses Schwinger's parametrization to raise the denominator factors into
exponential factors. The tensor integrals are dealt with by using differential
representations of the numerator factors. After doing as many loop 
integrations analytically as possible one ends up with four--fold parameter
integrations for the derivative of the normalization factor and the 
transition form factors which are evaluated numerically.

\section{Heavy quark spin symmetry}
\subsection{Structure of weak transitions in the HQL}
In the HQL the spins of the double--heavy diquark and the light quark in a 
double--heavy baryon decouple. At zero recoil and close to zero recoil this 
leads to spin symmetry relations among
transition form factors between double--heavy baryons and a zero recoil
normalization for the form factors.
The spin symmetry is exact at zero recoil and close to zero recoil where
the near--zero recoil region is specified later on.
Near zero recoil one can
choose the momenta as $p^\mu_1=m_{bc}v^\mu$ and
 $p^\mu_2=m_{cc}v'^\mu=m_{cc}v^\mu+r^\mu$ where $r$ is a small
residual momentum in the sense that $r^2\sim O(1)$ when $m_c\to\infty$.
Since the final baryon is on mass--shell one has
$v\cdot r= - r^2/2m_{cc}\sim O(1/m_c)$. This imposes a restriction
on the kinematical variable $w=v\cdot v'$ since 
$r^2=m^2_{cc}(v-v')^2=2m^2_{cc}(1-w)\sim O(1)$.
From the last equation one obtains $w\sim 1+O(1/m_c^2)$.
This situation is different compared with the case of baryons
with a single heavy quark. A baryon with a single heavy quark
possesses both a spin and a flavor symmetry. 
In what follows we will work near zero recoil in the sense that we 
neglect terms of  $O(v\cdot r)$. In order to keep things simple we assume that 
$m_{bc}=m_b+m_c$ and $m_{cc}=2m_c$. 
 
Using the above assumptions the heavy mass propagators simplify in the
HQL. One has   
\seq 
\eq 
\tilde S_b(k_1+p-\tfrac12 p') &\rightarrow& 
\frac{1+\not\! v}{2}\frac{1}{-k_1v - i \epsilon},
\\
&&\nonumber\\
\tilde S_c(k_1\pm\tfrac12 p') &\rightarrow& 
\frac{1\pm \not\! v'}{2}\frac{1}{\mp k_1v - i \epsilon} \,. 
\en 
\sen 
Due to the simple form of the heavy quark propagators in the HQL the
defining equation for the coupling constant Eq.(\ref{massop}) 
simplifies considerably. One obtains 
\eq\label{gXi} 
1 &=& N_{Q_1Q_2}\frac{g^2_{\,\Xi_{Q_1Q_2}}}{(16\pi^2)^2}
{\rm tr}\Big[\Gamma^{(d)}\frac{1+\not\! v}{2}
\bar\Gamma^{(d)} \frac{1-\not\! v}{2}\Big] 
\int\!\frac{d^4k_1}{\pi^2 i}\int\!\frac{d^4k_2}{(\pi^2 i}\,\,
\overline\Phi^2_B(y_0) 
\ \frac{\Gamma^{(q)}\tilde S_3(k_2-k_1){\bar \Gamma}^{(q)}}
{(- k_1v - i\epsilon)^2 (k_2v - i\epsilon)}\,.
\en 
A similar simplification occurs for the transition operator 
$\Lambda^\mu(p,p')$ in Eq.(\ref{transition}). One obtains
\eq 
\Lambda^\mu(v,v') &=& 
12\,\frac{ g_{\,\Xi_{bc}}\,g_{\,\Xi_{cc}} }{(16\pi^2)^2}\,
\ {\rm tr}\Big[ 
\Gamma^{(d)}_{cc}\frac{1+\not\! v'}{2}\,O^\mu\,
\frac{1+\not\! v}{2}\bar\Gamma^{(d)}_{bc}
\frac{1-\not\! v'}{2}\Big]
\nonumber\\
&\times& 
\int\frac{d^4k_1}{\pi^2 i}\int\frac{d^4k_2}{\pi^2 i} \, 
\overline\Phi_B(y_0) \, \overline\Phi_B(y_r) \, 
\frac{\Gamma^{(q)}_{cc}\tilde S_3(k_2-k_1) \bar\Gamma^{(q)}_{bc}}
{(- k_1v - i\epsilon) (- k_1v' - i\epsilon)(k_2v' - i\epsilon)}  
\label{HQL}
\en  
where 
\eq 
y_r \equiv y(r) =  - \tfrac12 (k_1-k_2)^2 
                   - \tfrac16 (k_1+k_2 - r)^2 \,. 
\en   
It is not difficult to see that once we put $v=v'\,\, (r=0)$
in Eq.~(\ref{HQL}) the two--loop integral and the trace factor in~(\ref{HQL})
reduce to the corresponding factors in Eq.~(\ref{gXi}).
As a result all semileptonic transition matrix elements
can be expressed in terms of a universal function 
$\eta(\omega)$ normalized to 1 at zero recoil with $\omega=1$ 
in full consistency with the heavy quark 
spin symmetry results derived in~\cite{Flynn:2007qt}.

Neglecting terms of $O(v\cdot r)$ one finds

\seq\label{sl_nzl} 
\eq 
\Lambda^\mu\Big(\Xi_{bc}(v)\to\Xi_{cc}(v')\Big)  &=& 
\sqrt{2}\, \Big(\gamma^\mu - \frac{2}{3} \gamma^\mu\gamma^5 \Big) 
\, \eta(w)\,,\\
&&\nonumber\\
\Lambda^\mu\Big(\Xi'_{bc}(v)\to\Xi_{cc}(v')\Big)  &=& 
- \sqrt{\tfrac{2}{3}}\, \gamma^\mu \gamma^5 \, \eta(w)\, \\
&&\nonumber\\
\Lambda^\mu\Big(\Xi_{bc}(v)\to\Xi^\ast_{cc}(v',\nu)\Big)  &=&  
- \sqrt{\tfrac{2}{3}}\, g^{\mu\nu} \,\eta(w),
\\
&&\nonumber\\
\Lambda^\mu\Big(\Xi'_{bc}(v)\to\Xi^\ast_{cc}(v',\nu)\Big)  &=& 
\sqrt{2} \, g^{\mu\nu} \, \eta(w),
\\
&&\nonumber\\ 
\Lambda^\mu\Big(\Xi^\ast_{bc}(v,\nu)\to\Xi_{cc}(v')\Big)  &=&   
- \sqrt{\tfrac{2}{3}}\, g^{\mu\nu} \, \eta(w),
\\
&&\nonumber\\
\Lambda^\mu\Big(\Xi^\ast_{bc}(v,\nu)\to\Xi^\ast_{cc}(v',\nu')\Big)  &=& 
\sqrt{2}\, O^\mu \, g^{\nu\nu'} \, \eta(w) \,.
\en 
\sen 
where 
\eq 
\label{eta}
\eta(w) &=& \frac{J(w)}{J(1)} \,, \\ 
J(w)  &=& \int\frac{d^4k_1}{\pi^2 i}\int\frac{d^4k_2}{\pi^2 i} \, 
\overline\Phi_B(y_0)  \, \overline\Phi_B(y_r)  \, 
\frac{m_q+(k_2-k_1)v } 
{(- k_1v - i\epsilon)^2 (k_2v - i\epsilon)
(m_q^2 - (k_2 - k_1)^2 - i\epsilon)}  \,.
\label{phi}
\en  
By keeping the corrections of $O(v\cdot r)$ one can obtain explicit 
model dependent expressions for the corrections
to the spin symmetry relations Eq.(\ref{sl_nzl}). These corrections will not 
be listed in the present paper.

As an extra bonus of our dynamical treatment the $bb \to bc$ transition 
matrix element can be related to the $bc \to cc$ transition in the HQL. 
One simply has to replace $m_{cc}$ by $m_{bb}$ in the functional form 
of the universal function $\eta(w)$ in (\ref{eta}), i.e. in the functions
$\overline\Phi_B(y_0)$ and $\overline\Phi_B(y_r)$ appearing in (\ref{phi}).

\subsection{Calculation of the universal function $\eta(w)$} 

It turns out that one can derive a closed--form expression for the universal
Isgur--Wise (IW) function $\eta(w)$ if one uses a Gaussian ansatz for the 
three--quark correlation function $\overline\Phi_B$ as has been done in
Eq.~(\ref{corr_Fun}). 
We use the Laplace transformation 
\eq 
\overline\Phi_B(z) = \int\limits_0^\infty ds \, \Phi_B^L(s) \, e^{-sz}  
\en 
and the integral representation
\eq 
\exp\Big( - \frac{s_1 s_2}{s_1 + s_2} x \Big) = \frac{s_1 + s_2}{\pi} \  
\int\limits_{-\infty}^{\infty} dt_1 \int\limits_{-\infty}^{\infty} dt_2  
\ \exp\Big( - s_1 \ ((t_1+\sqrt{x})^2+t_2^2) - s_2 \ (t_1^2 + t_2^2) \Big) \,. 
\en 
In terms of the variable $w$ one obtains
\eq 
\label{jofw}
J(w) = - \frac{2}{\pi} \ \int\limits_0^\infty \int\limits_0^\infty 
\int\limits_0^\infty \frac{d\alpha_1 \alpha_1 d\alpha_2 d\alpha_3}{\Delta^2} 
\Big( m_q + \frac{\alpha_1 + \alpha_2}{4\Delta} \Big) 
\int\limits_{-\infty}^{\infty} dt_1 \int\limits_{-\infty}^\infty dt_2 \ 
\overline\Phi_B'(z_w) \ \overline\Phi_B(z_1) \,,  
\en 
where 
\eq 
& &\Delta = \frac{3}{4} + \alpha_3\,, \hspace*{.5cm} 
\overline\Phi_B'(z) = d\overline\Phi_B(z)/dz\,, \nonumber\\
& &z_w \ \equiv \ z(w) \ = \ \biggl(t_1 + m_{cc} \sqrt{\frac{w-1}{3}}\biggr)^2 
                         +  t_2^2 + \frac{2}{3} \biggl( \alpha_3 m_q^2 
         + \frac{(1+\alpha_3)(\alpha_1^2 + \alpha_1 \alpha_2 + \alpha_2^2) 
         + \alpha_1 \alpha_2 \alpha_3}{4\Delta} \biggr) \,, 
\en  
and where $m_{q}$ is the mass of the light quark in the DHB. The integral
(\ref{jofw}) can be evaluated in closed form using  
the Gaussian ansatz for the correlation function 
Eq.~(\ref{corr_Fun}).
One obtains 
a rather simple form for the universal $\eta(w)$ function 
\eq\label{etaw} 
\eta(w) =  \exp\Big( - 3 ( w - 1) \, \frac{m_{cc}^2}{\Lambda_B^2} \Big) \,. 
\en 
The dependence on the light quark masses has disappeared due to 
cancelation effects between the numerator  $J(w)$ 
and the denominator $J(1)$. 

For the slope $\rho^2$ of $\eta(w)$ defined by 
$\eta(w)=1-\rho^2(w-1)+ ...$\, 
one obtains 
\eq 
\label{slope}
\rho^2 = - \frac{d\eta(w)}{dw}\Big|_{w=1} = 3 \frac{m_{cc}^2}{\Lambda_B^2} \,. 
\en 
As mentioned before, the HQL results for the $bb \to bc$--transitions can be 
obtained by the replacement $m_{cc}\to m_{bb}$ in the IW function. 
Accordingly the slope of the IW 
function for the $bb \to bc$ transitions is obtained from (\ref{slope})
by the replacement $m_{cc}\to m_{bb}$, i.e. the slope increases 
by the factor $m_{bb}^{2}/ m_{cc}^{2}$ when going from the $bc \to cc$ case to
the $bb \to bc$ transition if one uses the same size parameter $\Lambda_{B}$ 
in both cases. One should stress that there exists a spin--flavour symmetry at 
zero recoil $w = 1$ giving $\eta(1) \equiv 1$, which means that the 
$bc \to cc$ transition is identical to the $bb \to bc$ one. Close to zero 
recoil there exists only spin symmetry, because the IW functions 
for $bc \to cc$ and $bb \to bc$ transitions explicitly contain the flavor 
factors $m_{cc}$ and $m_{bb}$, respectively. 

\section{Results} 

We now proceed to present our numerical results. We first present results 
on the semileptonic rates using finite heavy quark masses, i.e. we
do not take the HQL for the transition form factors. In the finite
mass case we present results for both the electron/muon mode
and the $\tau$--mode. We then present results using the zero recoil form 
factors in Eq.~(\ref{sl_nzl}) and the universal function
$\eta(\omega)$ of~Eq.(\ref{etaw}) , i.e. we extent the validity of the near 
zero recoil approach to the whole kinematic range 
$0 \leq q^{2} \leq (M_{1}-M_{2})^{2}$. 
Also we present estimates of the width using the nonrelativistic quark model,
in which, as described before, the wave functions have the same spin--flavor 
structure as our relativistic current considered in the nonrelativistic limit. 
Note, in the naive quark model we drop the $q^2$--dependence of the 
corresponding 
form factors and use only their values at $q^2=0$. 
We choose the Gaussian form Eq.~(\ref{corr_Fun}) for the correlation function 
of the double heavy baryons. Our results depend on 
the following set of parameters: the constituent quark masses and 
the size parameter $\Lambda_B$. The parameters have been taken from a fit to
the properties of light, single and double heavy 
baryons in previous analysises~\cite{Ivanov:1996pz}. One has 
\begin{equation}
\begin{array}{cccccc}
m_{u(d)} & m_s & m_c & m_b & \Lambda_B \\  \hline
$\ \ 0.42\ \ $ & $\ \ 0.57\ \ $ & $\ \ 1.7\ \ $ & $\ \ 5.2\ \ $ 
& $\ \ 2.5 \ $ - $ \ 3.5\ \ $ & $\ \ {\rm GeV} $ \\
\end{array}\label{fitmas} \ \ \,. 
\end{equation} 
All our analytical calculations have been done using the computer program 
FORM~\cite{Vermaseren:2000nd}. For the numerical evaluation we have used 
FORTRAN. 

In Table 3 we present results on the $q^{2}=0$ values of the transition form
factors  $F_1^V$ and $F_1^A$ in the naive quark model. In Table 4 
we compare our finite mass results with the theoretical approaches of 
Ref.~\cite{Ebert:2004ck} and~\cite{Hernandez:2007qv} for the 
electron/muon mode. As indicated in Tables~4 and 5 we allow the size parameter
$\Lambda_B$ to vary in the range $2.5 \leq \Lambda_B \leq 3.5$~GeV. 
The variation
of our rates listed in Tables~4 and~5  reflect via error bars the variation 
of the size parameter $\Lambda_B$. Note that a smaller value of $\Lambda_B$ 
gives smaller rates and vice versa. In Table 5 we present detailed results on 
semileptonic rates of double heavy baryons in the exact finite mass approach 
for the electron/muon and tau--lepton modes. These are compared to the
corresponding results in the heavy quark 
spin symmetry limit and in the naive quark model. One can see that the naive 
quark model gives larger rates due to the omission of form factor effects 
that would result from the nonlocal structure of semileptonic transitions. 
The results in the heavy quark symmetry (HQS) limit are calculated using 
the IW function~(\ref{etaw}) with $m_{cc} = 2m_c = 3.4$ GeV for the 
$bc \to cc$ transitions and 
$m_{bb} = 2m_b = 10.4$~GeV for the $bb \to bc$ transitions. 

In order to check on the quality of the HQS limit in Figs.1 and 2 
we show plots of the $\omega$--dependence of the form factor 
$F_1^V(\omega)/\sqrt{2}$ for the exact finite mass result 
and the HQS result using the IW function $\eta(\omega)$. As examples we
take the two transitions $\Xi_{bc} \to \Xi_{cc}$ and $\Xi_{bb} \to \Xi_{bc}$. 
From Table 5 and Figs.1 and 2 one can see that HQS limit 
is better justified for $bb \to bc$ transitions, while there is a larger
difference between the finite mass results and the HQS results for the 
$bc \to cc$ mode as is evident in Fig.~1 and Table 5. 

\section{Summary} 

In this paper we have analyzed the semileptonic decay of double 
heavy baryons using a manifestly Lorentz covariant constituent quark model
approach. 
Our main results are summarized as follows:

-- We have derived results for the matrix elements of 
the semileptonic decays of double heavy baryons for finite 
values of the heavy quark/baryon masses and for the HQS limit which is valid
at and close to zero recoil;
 
-- We have presented a detailed numerical analysis of the decay rates 
for the two $(e/\mu)$-- and $\tau$--lepton modes in 
the exact finite mass approach and in the HQS limit; 

-- We have compared our results with the predictions of other theoretical 
approaches. 

We hope that the results of this paper can be used to extract 
the value of the CKM matrix element $V_{cb}$ from future experiments on 
the semileptonic decays of double heavy baryons. 

\begin{acknowledgments}

This work was supported by the DFG under Contract No. FA67/31-1,
No. FA67/31-2, and No. GRK683. M.A.I. appreciates the partial support
of the Heisenberg-Landau program and DFG grant KO 1069/12-1. 
This research is also part of the European Community-Research 
Infrastructure Integrating Activity ``Study of Strongly Interacting 
Matter'' (HadronPhysics2, Grant Agreement No. 227431) and of 
the President grant of Russia ``Scientific Schools''  No. 871.2008.2.

\end{acknowledgments}

\appendix\section{Nonrelativistic quark model: spin--flavor wave functions and 
semileptonic decay constants of double heavy baryons} 

In this Appendix we present results on the $q^{2}=0$ values of the double
heavy transition form factors $F_1^V(0)$ and $F_1^A(0)$  
in the nonrelativistic quark model. 
As emphasized before 
the nonrelativistic quark model is based on the spin--flavor wave functions 
which arise in the nonrelativistic limit of the relativistically covariant 
double heavy three--quark currents with quantum numbers $J^{P}=\frac{1}{2}^+$ 
and $\frac{3}{2}^+$. The corresponding quark model 
spin--flavor wave functions are given in Table 2, where we use the following
notation for the antisymmetric 
$\chi_A$ and symmetric ($\chi_S$, $\chi_S^\ast$)  
spin wave functions with $S_z=+1/2$ (spin projection on z--axis): 
\eq\label{spin_wf} 
\chi_A \, = \, \sqrt{\frac{1}{2}} \ \biggl\{ 
\uparrow (\uparrow \downarrow - \downarrow \uparrow) \biggr\}\,, 
\quad\quad \chi_S \, = \, \sqrt{\frac{1}{6}}  \ \biggl\{ 
\uparrow (\uparrow \downarrow + \downarrow \uparrow)  
- 2 \downarrow \uparrow \uparrow \biggr\} \,, 
\quad\quad \chi_S^\ast \, = \, \sqrt{\frac{1}{3}}  \ \biggl\{ 
  \uparrow \uparrow \downarrow + \uparrow \downarrow \uparrow   
+ \downarrow \uparrow \uparrow \biggr\} \,. 
\en 
We derive the expressions for the semileptonic decay constants 
$F_1^V(0)$ and $F_1^A(0)$ of double heavy baryons using the master formulas: 
\begin{eqnarray}
F_1^V(0) = <B^\prime|\sum\limits_{i=1}^3 [I_{bc}]^{(i)} |B>
\hspace*{.3cm} \mbox{and} \hspace*{.3cm}
F_1^A(0) = <B^\prime|\sum\limits_{i=1}^3 [\sigma_3 I_{bc}]^{(i)} |B>\,, 
\nonumber
\end{eqnarray}
where $\sigma_3$ is the $z$--component of the Pauli spin matrix and $I_{bc}$ 
is the flavor matrix responsible for the $b \to c$ semileptonic transitions. 
In Table 3 we list the results for the $\Xi_{Q_{1}Q_{2}}$--type baryon decay 
modes. Corresponding results for the $\Omega_{Q_{1}Q_{2}}$--type baryon decay 
modes can be obtained by replacing the light nonstrange quark $q$ by the 
strange quark $s$. 

\section{Spin--kinematics of semileptonic decays} 

In this Appendix we first write down covariant expressions for the 
current--induced transitions involving the $(1/2^{+})$ and $(3/2^{+})$
baryons which allows us to define sets of vector and axial vector invariant 
transition form factors. We then calculate all helicity 
amplitudes expressed in terms of linear combinations of the invariant form 
factors. The advantage
of using helicity amplitudes is that one obtains very compact expressions
for the decay rates including lepton mass effects. We mention that the use
of helicity amplitudes allows one also to derive very compact expressions
for single angular decay distributions and for joint angular decay 
distributions of the decay products
(see e.g. \cite{Bialas:1992ny,Kadeer:2005aq}).

The momenta and masses in the semileptonic decays of double--heavy 
baryons are denoted by 
\eq
B_1(p_1,M_1) \to B_2(p_1,M_2) + l(p_l,m_l) + \nu_l(p_\nu,0) \, 
\en 
where $p_1=p_2+q$ and $q=p_l+p_\nu$. 
The matrix elements of the vector and axial vector currents 
$J_\mu^{V(A)}$ between the baryon states with spin 1/2 or 3/2 
are written as 

Transition $\frac{1}{2}^+ \to \frac{1}{2}^+$\,: 
\seq
\eq 
& &M_\mu^V = \la B_2 | J_\mu^V | B_1 \ra = \bar u(p_2,s_2) 
\Big[ \gamma_\mu F_1^V(q^2) 
    - i \sigma_{\mu\nu} \frac{q_\nu}{M_1} F_2^V(q^2)  
    + \frac{q_\mu}{M_1} F_3^V(q^2) 
\Big] u(p_1,s_1) \\
& &M_\mu^A = \la B_2 | J_\mu^A | B_1 \ra = \bar u(p_2,s_2) 
\Big[ \gamma_\mu F_1^A(q^2)  
    - i \sigma_{\mu\nu} \frac{q_\nu}{M_1} F_2^A(q^2)   
    + \frac{q_\mu}{M_1} F_3^A(q^2)  
\Big] \gamma_5 u(p_1,s_1) 
\en 
\sen 

Transition $\frac{1}{2}^+ \to \frac{3}{2}^+$\,:
\seq 
\eq 
M_\mu^V &\!\! = \!\! & 
\la B_2^\ast | J_\mu^V | B_1 \ra = \bar u^\alpha(p_2,s_2) 
\Big[ g_{\alpha\mu} F_1^V(q^2)  
    + \gamma_\mu \frac{p_{1\alpha}}{M_1} F_2^V(q^2) \nonumber\\   
&+& \frac{p_{1\alpha} p_{2\mu}}{M_1^2} F_3^V(q^2)   
    + \frac{p_{1\alpha} q_\mu}{M_1^2}    F_4^V(q^2)       
\Big] \gamma_5 u(p_1,s_1) \\
M_\mu^A &\!\! = \!\! & 
\la B_2^\ast | J_\mu^A | B_1 \ra = \bar u^\alpha(p_2,s_2) 
\Big[ g_{\alpha\mu} F_1^A(q^2)  
    + \gamma_\mu \frac{p_{1\alpha}}{M_1} F_2^A(q^2) \nonumber\\
&+& \frac{p_{1\alpha} p_{2\mu}}{M_1^2} F_3^A(q^2)   
    + \frac{p_{1\alpha} q_\mu}{M_1^2}    F_4^A(q^2)       
\Big] u(p_1,s_1) 
\en 
\sen 

Transition $\frac{3}{2}^+ \to \frac{1}{2}^+$\,: 
\seq 
\eq 
M_\mu^V &\!\! = \!\! & 
\la B_2 | J_\mu^V | B_1^\ast \ra = \bar u(p_2,s_2) 
\Big[ g_{\alpha\mu} F_1^V(q^2)  
    + \gamma_\mu \frac{p_{2\alpha}}{M_2} F_2^V(q^2) \nonumber\\   
&+& \frac{p_{2\alpha} p_{1\mu}}{M_2^2} F_3^V(q^2)   
    + \frac{p_{2\alpha} q_\mu}{M_2^2}    F_4^V(q^2)       
\Big] \gamma_5 u^\alpha(p_1,s_1) \\
M_\mu^A &\!\! = \!\! & 
\la B_2 | J_\mu^A | B_1^\ast \ra = \bar u(p_2,s_2) 
\Big[ g_{\alpha\mu} F_1^A(q^2)  
    + \gamma_\mu \frac{p_{2\alpha}}{M_2} F_2^A(q^2) \nonumber\\
&+& \frac{p_{2\alpha} p_{1\mu}}{M_2^2} F_3^A(q^2)   
    + \frac{p_{2\alpha} q_\mu}{M_2^2}    F_4^A(q^2)       
\Big] u^\alpha(p_1,s_1) 
\en 
\sen 

Transition $\frac{3}{2}^+ \to \frac{3}{2}^+$\,:
\seq 
\eq 
\hspace*{-1.5cm}
M_\mu^V &\!\! = \!\! & 
\la B_2^\ast | J_\mu^V | B_1^\ast \ra = \bar u^\alpha(p_2,s_2) 
\Big[ g_{\alpha\beta} \Big( 
    \gamma_\mu F_1^V(q^2) 
    - i \sigma_{\mu\nu} \frac{q_\nu}{M_1} F_2^V(q^2)  
    + \frac{q_\mu}{M_1} F_3^V(q^2) \Big) \nonumber\\
&+&\frac{q_\alpha q_\beta}{M_1^2} 
\Big(
    \gamma_\mu F_4^V(q^2) 
    - i \sigma_{\mu\nu} \frac{q_\nu}{M_1} F_5^V(q^2)  
    + \frac{q_\mu}{M_1} F_6^V(q^2) \Big) \nonumber\\
&+&\frac{g_{\alpha\mu} q_\beta - g_{\beta\mu} q_\alpha}{M_1} F_7^V(q^2)
\Big] u^\beta(p_1,s_1) \\
\hspace*{-1.5cm} 
M_\mu^A &\!\! = \!\! & 
\la B_2^\ast | J_\mu^A | B_1^\ast \ra = \bar u^\alpha(p_2,s_2) 
\Big[ g_{\alpha\beta} \Big( 
    \gamma_\mu F_1^A(q^2) 
    - i \sigma_{\mu\nu} \frac{q_\nu}{M_1} F_2^A(q^2)  
    + \frac{q_\mu}{M_1} F_3^A(q^2) \Big) \nonumber\\
&+&\frac{q_\alpha q_\beta}{M_1^2} 
\Big(
    \gamma_\mu F_4^A(q^2) 
    - i \sigma_{\mu\nu} \frac{q_\nu}{M_1} F_5^A(q^2)  
    + \frac{q_\mu}{M_1} F_6^A(q^2) \Big) \nonumber\\
&+&\frac{g_{\alpha\mu} q_\beta - g_{\beta\mu} q_\alpha}{M_1} F_7^A(q^2)
\Big] \gamma_5 u^\beta(p_1,s_1)  
\en
\sen 
where 
$\sigma_{\mu\nu} = (i/2) (\gamma_\mu \gamma_\nu - \gamma_\nu \gamma_\mu)$ 
and all $\gamma$ matrices are defined as in Bjorken--Drell. 

Next we express the vector and axial helicity amplitudes 
$H_{\lambda_2\lambda_W}^{V,A}$ in terms of the invariant form factors 
$F_i^{V,A}$, 
where $\lambda_W = t, \pm 1, 0$ and $\lambda_2 = \pm 1/2, \pm 3/2$ are 
the helicity components of the $W_{\rm off--shell}$ and the daughter baryon, 
respectively. Since lepton mass effects are taken into account in this 
paper we need to retain the temporal component ``$t$'' of the 
four--currents $J_\mu^{V,A}$. 
We need to calculate the expressions 
\eq 
H_{\lambda_2\lambda_W}^{V,A} = M_\mu^{V,A}(\lambda_2) 
\bar\epsilon^{\,\ast\mu}(\lambda_W) \,. 
\en 
Note that the helicity of the parent baryon ($\lambda_1$) is fixed by the
relation 
$\lambda_1 = \lambda_2 - \lambda_W$. We shall work in the rest frame of 
the parent baryon $B_1$ with the daughter baryon $B_2$ moving in the 
positive $z$-direction:
$p_1 = (M_1, \vec{\bf 0})$, $p_2 = (E_2, 0, 0, |{\bf p}_2|)$ and 
$q = (q_0, 0, 0, - |{\bf p}_2|)$, where  
$E_2 = Q_+/(2M_1)$, $|{\bf p}_2| = \sqrt{Q_+Q_-}/(2M_1)$, 
$q_0 = M_1 - E_2$ and $Q_\pm = (M_1 \pm M_2)^2 -q^2$. 

The $J=\frac{1}{2}$ baryon spinors are 
given by 
\eq 
\bar u_2\Big(p_2, \pm \frac{1}{2}\Big) &=& \sqrt{E_2 + M_2} 
\Big( \chi_\pm^\dagger, \frac{\mp |{\bf p}_2|}{E_2 + M_2}  
\chi_\pm^\dagger \Big)\,, \nonumber\\
u_1\Big(p_1, \pm \frac{1}{2}\Big) &=& \sqrt{2M_1} 
\left(
\begin{array}{l}
\chi_\pm \\
0 \\
\end{array}
\right)
\en 
where $\chi_+ = \left(
\begin{array}{l}
1 \\
0 \\
\end{array} \right)$ 
and $\chi_- = \left(
\begin{array}{l}
0 \\
1 \\
\end{array} \right)$ are two--component Pauli spinors. 

The $J=\frac{3}{2}$ baryon spinors are 
defined as 
\eq 
u_\mu(p,s^\ast) = \sum\limits_{\lambda,s} 
\la 1\lambda\frac{1}{2}s | \frac{3}{2} s^\ast \ra 
\epsilon_\mu(p,\lambda) u(p,s) 
\en  
where $\la 1\lambda\frac{1}{2}s | \frac{3}{2} s^\ast \ra$ is 
the projection matrix element of the spin $\frac{3}{2}$ onto spin 
$\frac{1}{2}$; $\epsilon_\mu(p,\lambda)$ is the polarization 
vector and $u(p,s)$ are the ususal $J=\frac{1}{2}$ spinors  
defined above. In particular, the $J=\frac{3}{2}$ spinors with 
helicities $\lambda=\pm3/2,\pm1/2$ read: 
\eq 
& &u_\mu\Big(p, \pm \frac{3}{2}\Big) 
= \epsilon_\mu(p, \pm 1) u\Big(p, \pm \frac{1}{2}\Big) \,, 
\nonumber\\
& &u_\mu(p, \pm \frac{1}{2}) 
= \sqrt{\frac{2}{3}} \epsilon_\mu(p, 0) 
u\Big(p, \pm \frac{1}{2}\Big) 
+ \sqrt{\frac{1}{3}} \epsilon_\mu(p, \pm 1) 
u\Big(p, \mp \frac{1}{2}\Big) 
\en
The polarization vectors corresponding to the parent and daughter 
$J=\frac{3}{2}$ baryons are given by: 
\eq 
& &\,\,\epsilon(p_1,0) = (0,0,0,1)\,, \hspace*{2.05cm}  
\epsilon(p_1,\pm 1) = \frac{1}{\sqrt{2}} (0,\mp 1,-i,0)\,, \nonumber\\
& &\epsilon^\ast(p_2,0) = \frac{1}{M_2} (|{\bf p}_2|,0,0,E_2)\,, 
\hspace*{.5cm}   
\epsilon^\ast(p_2,\pm 1) = \frac{1}{\sqrt{2}} (0,\mp 1,i,0)  \,. 
\en 
The polarization vectors of the $W_{\rm off-shell}$ are written as 
\eq 
\bar\epsilon^{\,\ast\mu}(t) &=& \frac{1}{\sqrt{q^2}} 
(q_0, 0, 0, - |{\bf p}_2|)\,, \nonumber\\
\bar\epsilon^{\,\ast\mu}(\pm 1) &=& \frac{1}{\sqrt{2}} 
(0, \pm 1, i, 0)\,, \\
\bar\epsilon^{\,\ast\mu}(0) &=& \frac{1}{\sqrt{q^2}} 
(|{\bf p}_2|, 0, 0, q_0) \nonumber\,.
\en 
They satisfy the conditions 
\eq 
q_\mu \bar\epsilon^{\,\ast\mu}(\pm 1,0) = 0\,, \hspace*{.5cm} 
q_\mu \bar\epsilon^{\,\ast\mu}(t) = \sqrt{q^2}\,. 
\en 
Using the above formulas for the spin wave functions with definite helicities
one can then calculate the helicity amplitudes 
$H_{\lambda_2\lambda_W} = H_{\lambda_2\lambda_W}^V 
- H_{\lambda_2\lambda_W}^A$, where the vector and axial components 
for the different spin transitions are defined by: 

Transition $\frac{1}{2}^+ \to \frac{1}{2}^+$\,:
\eq 
H_{\frac{1}{2}t}^V &=& \alpha_{\frac{1}{2}t}^V \, 
\Big( F_1^V M_- + F_3^V \frac{q^2}{M_1} \Big) \,,\nonumber\\[2mm]
H_{\frac{1}{2}0}^V &=& \alpha_{\frac{1}{2}0}^V \,
\Big( F_1^V M_+ + F_2^V \frac{q^2}{M_1} \Big) \,,\nonumber\\[2mm]
H_{\frac{1}{2}1}^V &=& \alpha_{\frac{1}{2}1}^V 
\Big( - F_1^V - F_2^V \frac{M_+}{M_1} \Big) \,,\nonumber\\[2mm]
& &\\[2mm]
H_{\frac{1}{2}t}^A &=& \alpha_{\frac{1}{2}t}^A \, 
\Big( F_1^A M_+ - F_3^A \frac{q^2}{M_1} \Big) \,,\nonumber\\[2mm]
H_{\frac{1}{2}0}^A &=& \alpha_{\frac{1}{2}0}^A \,   
\Big( F_1^A M_- - F_2^A \frac{q^2}{M_1} \Big) \,, \nonumber\\[2mm]
H_{\frac{1}{2}1}^A &=& \alpha_{\frac{1}{2}1}^A \,  
\Big( - F_1^A + F_2^A \frac{M_-}{M_1} \Big) \,,\nonumber 
\en 
where 
\eq 
M_\pm &=& M_1 \pm M_2 \,, \hspace*{3.8cm}
   q^2 = 2 M_1 M_2 (w_{\rm max} - w)\,, 
\nonumber\\[3mm]  
w &=& \frac{p_1p_2}{M_1M_2} \ = \ \frac{M_1^2 + M_2^2 - q^2}{2M_1M_2} \,, 
\hspace*{.5cm} w_{\rm max} \ = \ \frac{M_1^2 + M_2^2}{2M_1M_2} \,, 
\nonumber\\[3mm]   
\alpha_{\frac{1}{2}t}^V &=& \alpha_{\frac{1}{2}0}^A \ = \
\sqrt{\frac{2M_1M_2(w+1)}{q^2}} \,, \nonumber\\[2mm]   
\alpha_{\frac{1}{2}0}^V &=& \alpha_{\frac{1}{2}t}^A \ = \ 
\sqrt{\frac{2M_1M_2(w-1)}{q^2}} \,, \\[2mm]  
\alpha_{\frac{1}{2}1}^V &=&  
2 \sqrt{M_1M_2(w-1)} \,, \nonumber\\[4mm]  
\alpha_{\frac{1}{2}1}^A &=& 
2 \sqrt{M_1M_2(w+1)} \,. \nonumber
\en 
From parity or from explicit calculations one has 
for the $\frac{1}{2}^+ \to \frac{1}{2}^+$ helicity amplitudes 
\eq 
H^V_{-\lambda_2,-\lambda_W} = H^V_{\lambda_2,\lambda_W}\,, \hspace*{1cm} 
H^A_{-\lambda_2,-\lambda_W} = - H^A_{\lambda_2,\lambda_W}\,. 
\en 
Transition $\frac{1}{2}^+ \to \frac{3}{2}^+$\,: 
\eq 
H_{\frac{1}{2}t}^V &=& - \sqrt{\frac{2}{3}} \, \alpha_{\frac{1}{2}t}^V \, 
(w-1) \Big( F_1^V M_1 - F_2^V M_+ + F_3^V \frac{M_2}{M_1} (M_1 w - M_2)  
+ F_4^V \frac{q^2}{M_1} \Big) \,,\nonumber\\[2mm]
H_{\frac{1}{2}0}^V &=& - \sqrt{\frac{2}{3}} \, \alpha_{\frac{1}{2}0}^V \, 
\Big( F_1^V (M_1 w - M_2) - F_2^V (w+1) M_- + F_3^V (w^2 - 1) M_2 
\Big) \,,\nonumber\\[2mm]
H_{\frac{1}{2}1}^V &=& \frac{1}{\sqrt{6}} \, \alpha_{\frac{1}{2}1}^V \,  
\Big( F_1^V - 2 F_2^V (w+1) \Big) \,,\nonumber\\[2mm]
H_{\frac{3}{2}1}^V &=& - \frac{1}{\sqrt{2}} \, 
\alpha_{\frac{1}{2}1}^V \, F_1^V \,,\nonumber\\
& &\\[2mm] 
H_{\frac{1}{2}t}^A &=& \sqrt{\frac{2}{3}} \,  \alpha_{\frac{1}{2}t}^A \, 
(w+1) \Big( F_1^A M_1 + F_2^A M_- + F_3^A \frac{M_2}{M_1} (M_1 w - M_2)   
+ F_4^A \frac{q^2}{M_1}\Big) \,,\nonumber\\[2mm]
H_{\frac{1}{2}0}^A &=&  \sqrt{\frac{2}{3}} \alpha_{\frac{1}{2}0}^A 
\Big( F_1^A (M_1 w - M_2) + F_2^A (w-1) M_+ + F_3^A (w^2 - 1) M_2 
\Big) \,,\nonumber\\[2mm]
H_{\frac{1}{2}1}^A &=& \frac{1}{\sqrt{6}} \, \alpha_{\frac{1}{2}1}^A 
\Big( F_1^A - 2 F_2^A (w-1) \Big) \,,\nonumber\\
H_{\frac{3}{2}1}^A &=& \frac{1}{\sqrt{2}} \, \alpha_{\frac{1}{2}1}^A 
F_1^A \,.\nonumber
\en 
From parity or from explicit calculations one has 
for the $\frac{1}{2}^+ \to \frac{1}{3}^+$ helicity amplitudes 
\eq 
H^V_{-\lambda_2,-\lambda_W} = - H^V_{\lambda_2,\lambda_W}\,, \hspace*{1cm} 
H^A_{-\lambda_2,-\lambda_W} =   H^A_{\lambda_2,\lambda_W}\,. 
\en 
Transition $\frac{3}{2}^+ \to \frac{1}{2}^+$\,:
\eq 
H_{\frac{1}{2}t}^V &=& 
- \sqrt{\frac{2}{3}} \, \alpha_{\frac{1}{2}t}^V \, (w - 1) 
\Big( F_1^V M_2 + F_2^V M_+ - F_3^V \frac{M_1}{M_2} (M_1 - M_2 w) 
- F_4^V \frac{q^2}{M_2}\Big) \,,\nonumber\\[2mm]
H_{\frac{1}{2}0}^V &=& - \sqrt{\frac{2}{3}} \, \alpha_{\frac{1}{2}0}^V \, 
\Big( F_1^V (M_1 - M_2 w) + F_2^V (w+1) M_- - F_3^V (w^2 - 1) M_1  
\Big) \,,\nonumber\\[2mm]
H_{\frac{1}{2}1}^V &=& \frac{1}{\sqrt{6}} \, \alpha_{\frac{1}{2}1}^V \, 
\Big( F_1^V + 2 F_2^V (w+1) \Big) \,,\nonumber\\
H_{\frac{1}{2}-1}^V &=& \frac{1}{\sqrt{2}} \alpha_{\frac{1}{2}1}^V \,  
F_1^V \,,\nonumber\\[2mm] 
& &\\[2mm] 
H_{\frac{1}{2}t}^A &=& \sqrt{\frac{2}{3}} \, \alpha_{\frac{1}{2}t}^A \, 
(w+1) \, \Big( F_1^A M_2 - F_2^A M_- - F_3^A \frac{M_1}{M_2} (M_1 - M_2 w) \, 
- F_4^A \frac{q^2}{M_2}\Big) \,,\nonumber\\[2mm]
H_{\frac{1}{2}0}^A &=& - \sqrt{\frac{2}{3}} \, \alpha_{\frac{1}{2}0}^A \, 
\Big( - F_1^A (M_1 - M_2 w) + F_2^A (w-1) M_+ + F_3^A (w^2 - 1) M_1  
\Big) \,,\nonumber\\[2mm]
H_{\frac{1}{2}1}^A &=& \frac{1}{\sqrt{6}} \, \alpha_{\frac{1}{2}1}^A \, 
\Big( - F_1^A + 2 F_2^A (w-1) \Big) \,,\nonumber\\[2mm]
H_{\frac{3}{2}1}^A &=& - \frac{1}{\sqrt{2}} \, \alpha_{\frac{1}{2}1}^A \, 
F_1^A \,,\nonumber
\en 
From parity or from explicit calculations one has 
for the $\frac{3}{2}^+ \to \frac{1}{2}^+$ helicity amplitudes 
\eq 
H^V_{-\lambda_2,-\lambda_W} = - H^V_{\lambda_2,\lambda_W}\,, \hspace*{1cm}  
H^A_{-\lambda_2,-\lambda_W} =   H^A_{\lambda_2,\lambda_W}\,. 
\en 
Transition $\frac{3}{2}^+ \to \frac{3}{2}^+$\,: 
\eq 
H_{\frac{1}{2}t}^V &=& - \frac{1+2w}{3} H_{\frac{1}{2}t}^V(F_1^V,F_2^V,F_3^V) 
+ \frac{2}{3} (w^2-1) \frac{M_2}{M_1} H_{\frac{1}{2}t}^V(F_4^V,F_5^V,F_6^V) 
\,,\nonumber\\[2mm]
H_{\frac{1}{2}0}^V &=& - \frac{1+2w}{3} H_{\frac{1}{2}0}^V(F_1^V,F_2^V,F_3^V) 
+ \frac{2}{3} (w^2-1) \frac{M_2}{M_1} H_{\frac{1}{2}0}^V(F_4^V,F_5^V,F_6^V) 
- \frac{1}{3} \, \alpha_{\frac{1}{2}1}^A \, (w+1) \,  
\frac{\sqrt{2q^2}}{M_1} \, F_7^A \,,\nonumber\\[2mm]
H_{\frac{1}{2}1}^V &=& - \frac{2w}{3} H_{\frac{1}{2}1}^V(F_1^V,F_2^V,F_3^V) 
+ \frac{2}{3} (w^2-1) \frac{M_2}{M_1} H_{\frac{1}{2}1}^V(F_4^V,F_5^V,F_6^V) 
+ \frac{1}{3} \alpha_{\frac{1}{2}1}^V (w+1) \frac{M_+}{M_1} F_7^V
\,,\nonumber\\[2mm]
H_{\frac{3}{2}1}^V &=& - \frac{1}{\sqrt{3}} 
H_{\frac{1}{2}1}^V(F_1^V,F_2^V,F_3^V) 
+ \frac{1}{\sqrt{3}} \alpha_{\frac{1}{2}1}^V (w+1) \frac{M_2}{M_1} F_7^V
\,,\nonumber\\[2mm]
H_{\frac{1}{2}-1}^V &=& - \frac{1}{\sqrt{3}} 
H_{\frac{1}{2}1}^V(F_1^V,F_2^V,F_3^V) 
+ \frac{1}{\sqrt{3}} \alpha_{\frac{1}{2}1}^V (w+1) F_7^V
\,,\nonumber\\[2mm]
H_{\frac{3}{2}t}^V &=& - H_{\frac{1}{2}t}^V(F_1^V,F_2^V,F_3^V) 
\,,\nonumber\\[2mm]
H_{\frac{3}{2}0}^V &=& - H_{\frac{1}{2}0}^V(F_1^V,F_2^V,F_3^V) 
\,,\nonumber\\[2mm]
& &\\[2mm] 
H_{\frac{1}{2}t}^A &=& \frac{1-2w}{3} H_{\frac{1}{2}t}^A(F_1^V,F_2^V,F_3^V) 
+ \frac{2}{3} (w^2-1) \frac{M_2}{M_1} H_{\frac{1}{2}t}^A(F_4^A,F_5^A,F_6^A) 
\,,\nonumber\\[2mm]
H_{\frac{1}{2}0}^A &=& \frac{1-2w}{3} H_{\frac{1}{2}0}^A(F_1^A,F_2^A,F_3^A) 
+ \frac{2}{3} (w^2-1) \frac{M_2}{M_1} H_{\frac{1}{2}0}^A(F_4^A,F_5^A,F_6^A) 
+ \frac{1}{3} \, \alpha_{\frac{1}{2}1}^A \, (w-1) \,  
\frac{\sqrt{2q^2}}{M_1} \, F_7^A \,,\nonumber\\[2mm] 
H_{\frac{1}{2}1}^A &=& - \frac{2w}{3} H_{\frac{1}{2}1}^A(F_1^A,F_2^A,F_3^A) 
+ \frac{2}{3} (w^2-1) \frac{M_2}{M_1} H_{\frac{1}{2}1}^A(F_4^A,F_5^A,F_6^A) 
- \frac{1}{3} \alpha_{\frac{1}{2}1}^A (w-1) \frac{M_-}{M_1} F_7^A 
\,,\nonumber\\[2mm]
H_{\frac{3}{2}1}^A &=& - \frac{1}{\sqrt{3}} 
H_{\frac{1}{2}1}^A(F_1^A,F_2^A,F_3^A) 
- \frac{1}{\sqrt{3}} \alpha_{\frac{1}{2}1}^A 
(w-1) \frac{M_2}{M_1} F_7^A \,,\nonumber\\[2mm]
H_{\frac{1}{2}-1}^A &=& \frac{1}{\sqrt{3}} 
H_{\frac{1}{2}1}^A(F_1^A,F_2^A,F_3^A) - \frac{1}{\sqrt{3}} 
\alpha_{\frac{1}{2}1}^A (w-1) F_7^A 
\,,\nonumber\\[2mm]
H_{\frac{3}{2}t}^A &=& - H_{\frac{1}{2}t}^A(F_1^A,F_2^A,F_3^A) 
\,,\nonumber\\[2mm] 
H_{\frac{3}{2}0}^A &=& - H_{\frac{1}{2}0}^A(F_1^A,F_2^A,F_3^A) \,,\nonumber
\en 
where 
\eq 
H_{\frac{1}{2}t}^V(x,y,z) &=& \alpha_{\frac{1}{2}t}^V  
\Big( x M_- + z \frac{q^2}{M_1} \Big) \,,\nonumber\\[2mm] 
H_{\frac{1}{2}0}^V(x,y,z) &=& \alpha_{\frac{1}{2}0}^V  
\Big( x M_+ + y \frac{q^2}{M_1} \Big) \,,\nonumber\\[2mm]
H_{\frac{1}{2}1}^V(x,y,z) &=& - \alpha_{\frac{1}{2}1}^V   
\Big( x + y \frac{M_+}{M_1} \Big) \,,\nonumber\\[2mm]
& &\\
H_{\frac{1}{2}t}^A(x,y,z) &=& \alpha_{\frac{1}{2}t}^A   
\Big( x M_+ - z \frac{q^2}{M_1} \Big) \,,\nonumber\\[2mm]
H_{\frac{1}{2}0}^A(x,y,z) &=& \alpha_{\frac{1}{2}0}^A    
\Big( x M_- - y \frac{q^2}{M_1} \Big) \,,\nonumber\\[2mm]
H_{\frac{1}{2}1}^A(x,y,z) &=& \alpha_{\frac{1}{2}1}^A    
\Big( - x + y \frac{M_-}{M_1} \Big) \,,\nonumber 
\en 
From parity or from explicit calculations one has 
for the $\frac{3}{2}^+ \to \frac{3}{2}^+$ helicity amplitudes 
\eq 
H^V_{-\lambda_2,-\lambda_W} =   H^V_{\lambda_2,\lambda_W}\,, \hspace*{1cm} 
H^A_{-\lambda_2,-\lambda_W} = - H^A_{\lambda_2,\lambda_W}\,. 
\en 
The decay width is given by the expression 
\eq 
\Gamma_{s_1 \to s_2} = N_{s_1s_2} 
\frac{G_F^2 |V_{\rm CKM}|^2}{192 \pi^3} \ 
\frac{M_2}{M_1^2} \ \int\limits_{m_l^2}^{M_-^2} 
\frac{dq^2}{q^2} \ (q^2 - m_l^2)^2 \sqrt{w^2-1} \ {\cal H}_{s_1 \to s_2}     
\en 
where $N_{s_1s_2} = 1$ for $s_1 = 1/2$ and 
$1/2$ for $s_1 = 3/2$; 
${\cal H}_{s_1 \to s_2}$ are the bilinear combinations of 
the helicity amplitudes: 
\seq 
\eq 
{\cal H}_{\frac{1}{2} \to \frac{1}{2}} &=&  
  |H_{\frac{1}{2}1}|^2 +  |H_{-\frac{1}{2}-1}|^2 
+ |H_{\frac{1}{2}0}|^2 +  |H_{-\frac{1}{2}0}|^2 \nonumber\\
&+&\frac{m_l^2}{2q^2} \Big( 
3 |H_{\frac{1}{2}t}|^2 + 3 |H_{-\frac{1}{2}t}|^2 
+ |H_{\frac{1}{2}1}|^2 +   |H_{-\frac{1}{2}-1}|^2  
+ |H_{\frac{1}{2}0}|^2 +   |H_{-\frac{1}{2}0}|^2 
\Big) \\[3mm]
{\cal H}_{\frac{1}{2} \to \frac{3}{2}} &=&  
   |H_{\frac{1}{2}1}|^2 + |H_{-\frac{1}{2}-1}|^2 
 + |H_{\frac{3}{2}1}|^2 + |H_{-\frac{3}{2}-1}|^2   
 + |H_{\frac{1}{2}0}|^2 + |H_{-\frac{1}{2}0}|^2 \nonumber\\
&+&\frac{m_l^2}{2q^2} 
  \Big(3 |H_{\frac{1}{2}t}|^2 + 3 |H_{-\frac{1}{2}t}|^2 
 + |H_{\frac{1}{2}1}|^2 + |H_{-\frac{1}{2}-1}|^2  
 + |H_{\frac{3}{2}1}|^2 + |H_{-\frac{3}{2}-1}|^2  \nonumber\\
&+&|H_{\frac{1}{2}0}|^2 + |H_{-\frac{1}{2}0}|^2 
\Big)\\[3mm] 
{\cal H}_{\frac{3}{2} \to \frac{1}{2}} &=&  
|H_{\frac{1}{2}1}|^2  + |H_{-\frac{1}{2}-1}|^2 +  
|H_{\frac{1}{2}-1}|^2 + |H_{-\frac{1}{2}1}|^2 +  
|H_{\frac{1}{2}0}|^2  + |H_{-\frac{1}{2}0}|^2 \nonumber\\
&+&\frac{m_l^2}{2q^2} 
    \Big(3 |H_{\frac{1}{2}t}|^2 + 3 |H_{-\frac{1}{2}t}|^2 
 +  |H_{\frac{1}{2}1}|^2  + |H_{-\frac{1}{2}-1}|^2  
 +  |H_{\frac{1}{2}-1}|^2 + |H_{-\frac{1}{2}1}|^2  \nonumber\\
&+& |H_{\frac{1}{2}0}|^2  + |H_{-\frac{1}{2}0}|^2 
\Big)\\[3mm] 
{\cal H}_{\frac{3}{2} \to \frac{3}{2}} &=&  
    |H_{\frac{1}{2}1}|^2  + |H_{-\frac{1}{2}-1}|^2 
  + |H_{\frac{3}{2}1}|^2  + |H_{-\frac{3}{2}-1}|^2   
  + |H_{\frac{1}{2}-1}|^2 + |H_{-\frac{1}{2}1}|^2 \nonumber\\ 
&+& |H_{\frac{1}{2}0}|^2  + |H_{-\frac{1}{2}0}|^2 
  + |H_{\frac{3}{2}0}|^2  + |H_{-\frac{3}{2}0}|^2 
  + \frac{m_l^2}{2q^2} 
    \Big(3 |H_{\frac{1}{2}t}|^2 + 3 |H_{-\frac{1}{2}t}|^2\nonumber\\ 
&+& 3|H_{\frac{3}{2}t}|^2 + 3 |H_{-\frac{3}{2}t}|^2 
 +  |H_{\frac{1}{2}1}|^2 + |H_{-\frac{1}{2}-1}|^2 
 +  |H_{\frac{3}{2}1}|^2 + |H_{-\frac{3}{2}-1}|^2  \nonumber\\ 
&+& |H_{\frac{3}{2}0}|^2  + |H_{-\frac{3}{2}0}|^2  
 +  |H_{\frac{1}{2}-1}|^2 + |H_{-\frac{1}{2}1}|^2  
 +  |H_{\frac{1}{2}0}|^2  + |H_{-\frac{1}{2}0}|^2 
\Big)  
\en 
\sen 

In the zero recoil limit the expressions for the rates simplify considerably.  
One obtains 
\seq 
\eq 
\Gamma_{\frac{1}{2} \to \frac{1}{2}} &=& \beta_W 
\int\limits_{1}^{w_{\rm max}} 
dw \ \sqrt{w^2-1} \Big(l_1^+(w) F^2(w) + l_1^-(w) G^2(w)\Big)\,, \\ 
\Gamma_{\frac{1}{2} \to \frac{3}{2}} &=& \beta_W  
\int\limits_{1}^{w_{\rm max}} dw \sqrt{w^2-1} 
\ l_2(w) \ G^2(w)\,, \\ 
\Gamma_{\frac{3}{2} \to \frac{1}{2}} &=& \frac{1}{2} \beta_W  
\int\limits_{1}^{w_{\rm max}} dw \sqrt{w^2-1} 
\ l_3(w) \ G^2(w)\,, \\ 
\Gamma_{\frac{3}{2} \to \frac{3}{2}} &=& \frac{1}{2} \beta_W  
\int\limits_{1}^{w_{\rm max}} dw \sqrt{w^2-1} 
\ \biggl( l_4^+(w) F^2(w) + l_4^-(w) G^2(w) \biggr)\,, 
\en 
\sen 
where $F = F_1^V$, $G = F_1^A$, $r = M_2/M_1$, $w = (1 + r^2)/(2r)$ and  
\eq 
\beta_W &=& \frac{G_F^2 |V_{\rm CKM}|^2}{12 \pi^3} \ M_1^5 \, r^4\,, 
\nonumber\\
l_1^\pm(w)  &=& (w \mp 1) (3 w_{\rm max} \pm 1 - 2 w) \,, \nonumber\\  
l_2(w)      &=&  2 (w + 1) \biggl( w_{\rm max} - w 
+ \frac{w^2 - 1}{6 r} \biggr) \,, \nonumber\\  
l_3(w)      &=&  2 (w + 1) \biggl( w_{\rm max} - w 
+ \frac{(w^2 - 1) r}{6} \biggr) \,, \nonumber\\  
l_4^\pm(w)  &=& \frac{4}{9} \biggl( l_1^\pm(w) \frac{3 + 2w^2}{2} 
           \pm (w_{\rm max} \pm 1) (w^2 - 1) \biggr)  \, . 
\en

\newpage 

\begin{table}
\begin{center}
{\bf Table 1.} 
Classification and mass values of double--heavy baryons. \\ 
Mass values are used from \cite{Ebert:2004ck} except for 
the $\Xi_{cc}$ mass which is taken from~\cite{Amsler:2008zzb}.  

\vspace*{.5cm}
\def\arraystretch{1.1}
\begin{tabular}{|c|c|c|c|c|}
\hline 
\,\, Notation \,\,  & \,\, Content \,\, &  \,\, $J^P$ \,\, &
\,\, $S_d$ \,\, & \,\, Mass (GeV)\,\, \\[2mm]
\hline
$\Xi_{cc}$  & $q\{cc\}$ & $1/2^+$ & $1$ & $3.5189$ \\
\hline
$\Xi_{bc}$  & $q\{bc\}$ & $1/2^+$ & $1$ & $6.933$ \\ 
\hline
$\Xi'_{bc}$ & $q[bc]$   & $1/2^+$ & $0$ & $6.963$ \\
\hline
$\Xi_{bb}$  & $q\{bb\}$ & $1/2^+$ & $1$ & $10.202$ \\
\hline
$\Xi_{cc}^\ast$  & $q\{cc\}$ & $3/2^+$ & $1$ & $3.727$ \\
\hline
$\Xi_{bc}^\ast$  & $q\{bc\}$ & $3/2^+$ & $1$ & $6.980$ \\
\hline
$\Xi_{bb}^\ast$  & $q\{bb\}$ & $3/2^+$ & $1$ & $10.237$ \\
\hline
$\Omega_{cc}$  & $s\{cc\}$ & $1/2^+$ & $1$ & $3.778$ \\
\hline
$\Omega_{bc}$  & $s\{bc\}$ & $1/2^+$ & $1$ & $7.088$ \\ 
\hline
$\Omega'_{bc}$ & $s[bc]$   & $1/2^+$ & $0$ & $7.116$ \\
\hline
$\Omega_{bb}$  & $s\{bb\}$ & $1/2^+$ & $1$ & $10.359$ \\
\hline
$\Omega_{cc}^\ast$  & $s\{cc\}$ & $3/2^+$ & $1$ & $3.872$ \\
\hline
$\Omega_{bc}^\ast$  & $s\{bc\}$ & $3/2^+$ & $1$ & $7.130$ \\
\hline
$\Omega_{bb}^\ast$  & $s\{bb\}$ & $3/2^+$ & $1$ & $10.389$ \\
\hline
\end{tabular}
\end{center}
\end{table}

\begin{table}
\begin{center} 
{\bf Table 2.} 
Double--heavy baryon wave functions. 

\vspace*{.5cm}
\def\arraystretch{1.5}
\begin{tabular}{|c|c||c|c|} 
\hline
  $\;\;$ Baryon $\;\;$ & $\;\;\;\;\;\;\;$ Wave function $\;\;\;\;\;\;\;$ &
  $\;\;$ Baryon $\;\;$ & $\;\;\;\;\;\;\;$ Wave function $\;\;\;\;\;\;\;$ \\ 
\hline 
$\Xi_{cc}$ & $qcc \,\,\, \chi_S$ &
$\Omega_{cc}$ & $scc \,\,\, \chi_S$ \\ 
\hline 
$\Xi_{bb}$ & $qbb \,\,\, \chi_S$ & 
$\Omega_{bb}$ & $sbb \,\,\, \chi_S$ \\ 
\hline 
$\Xi_{bc}$ & $\frac{1}{\sqrt{2}} q (bc + cb)  \,\,\, \chi_S$ & 
$\Omega_{bc}$ & $\frac{1}{\sqrt{2}} s (bc + cb) \,\,\, \chi_S$ \\
\hline 
$\Xi'_{bc}$ & $\frac{1}{\sqrt{2}} q (bc - cb)  \,\,\, \chi_A$ &
$\Omega'_{bc}$ & $\frac{1}{\sqrt{2}} s (bc - cb) \,\,\, \chi_A$ \\
\hline 
$\Xi^\ast_{cc}$ & - $qcc \,\,\, \chi_S^\ast$ &
$\Omega^\ast_{cc}$ & - $scc \,\,\, \chi_S^\ast$ \\
\hline 
$\Xi^\ast_{bb}$ & - $qbb \,\,\, \chi_S^\ast$ & 
$\Omega^\ast_{bb}$ & - $sbb \,\,\, \chi_S^\ast$ \\
\hline 
$\Xi^\ast_{bc}$ & - $\frac{1}{\sqrt{2}} q (bc+cb) \,\,\, \chi_S^\ast$ & 
$\Omega^\ast_{bc}$ & - $\frac{1}{\sqrt{2}} s (bc+cb) \,\,\, \chi_S^\ast$ \\
\hline
\end{tabular}
\end{center} 
\end{table}

\begin{table}
\begin{center} 
{\bf Table 3.} $F_1^V(0)$ and $G_1^V(0)$ 
in the nonrelativistic quark model. 
\vspace*{0.2cm}
\def\arraystretch{1.85}
\begin{tabular}{|c|c|c|c|c|c|c|}
\hline
Quantity & $\Xi_{bc}\to\Xi_{cc}$ & $\Xi'_{bc}\to\Xi_{cc}$ 
         & $\Xi_{bc}\to\Xi^\ast_{cc}$ & $\Xi'_{bc}\to\Xi^\ast_{cc}$  
         & $\Xi^\ast_{bc}\to\Xi_{cc}$ & $\Xi^\ast_{bc}\to\Xi^\ast_{cc}$ \\ 
         & $\Xi_{bb}\to\Xi_{bc}$ & $\Xi_{bb}\to\Xi'_{bc}$ 
         & $\Xi^\ast_{bb}\to\Xi_{bc}$ & $\Xi^\ast_{bb}\to\Xi'_{bc}$  
         & $\Xi_{bb}\to\Xi^\ast_{bc}$ & $\Xi^\ast_{bb}\to\Xi^\ast_{bc}$ \\
\hline
$F_1^V(0)$  & $\sqrt{2}$   
& $0$                  & $0$           
& $0$           & $0$   
& $\sqrt{2}$ \\
\hline
$F_1^A(0)$  
& $\frac{1}{\sqrt{2}}$  & $\sqrt{\frac{2}{3}}$ 
& $\frac{2}{3}$ & - $\frac{2}{\sqrt{3}}$  
& $\frac{\sqrt{2}}{3}$ 
& $\frac{\sqrt{2}}{3}$ \\
\hline
\end{tabular}
\end{center}
\end{table}

\begin{table}
\begin{center}
{\bf Table 4.} 
Semileptonic decay widths of double heavy baryons 
in units of $10^{-14}$ GeV. \\ 
Comparison with other approaches in case of 
light leptons $(e, \mu)$ in the final state. 

\vspace*{.5cm}
\def\arraystretch{1.1}
\begin{tabular}{|c|c|c|c||c|c|c|c|}
\hline 
   \,\, Decay mode \,\,  
&  \,\, Ref.~\cite{Ebert:2004ck}     \,\, 
&  \,\, Ref.~\cite{Hernandez:2007qv} \,\, 
&  \,\, Our results \,\, 
&   \,\, Decay mode \,\,  
&  \,\, Ref.~\cite{Ebert:2004ck}     \,\, 
&  \,\, Ref.~\cite{Hernandez:2007qv} \,\, 
&  \,\, Our results \,\, \\[2mm]
\hline
$\Xi_{bb} \to \Xi_{bc}$        
& 1.63 & $1.92^{+0.25}_{-0.05}$ & 0.80 $\pm$ 0.30 & 
$\Omega_{bb} \to \Omega_{bc}$  
& 1.70 & $2.14^{+0.20}_{-0.02}$ & 0.86 $\pm$ 0.32 
\\ 
\hline
$\Xi_{bc} \to \Xi_{cc}$        
& 2.30 & $2.57^{+0.26}_{-0.03}$ & 2.10 $\pm$ 0.70 &
$\Omega_{bc} \to \Omega_{cc}$  
& 2.48 & $2.59^{+0.20}$         & 1.88 $\pm$ 0.62 
\\ 
\hline
$\Xi'_{bc} \to \Xi_{cc}$       
& 0.88 & $1.36^{+0.10}_{-0.03}$ & 1.10 $\pm$ 0.32 & 
$\Omega'_{bc} \to \Omega_{cc}$ 
& 0.95 & $1.36^{+0.09}$         & 0.98 $\pm$ 0.28 
\\ 
\hline
$\Xi_{bb} \to \Xi'_{bc}$       
& 0.82 & $1.06^{+0.13}_{-0.03}$ & 0.43 $\pm$ 0.12 & 
$\Omega_{bb} \to \Omega'_{bc}$ 
& 0.83 & $1.16^{+0.13}$         & 0.48 $\pm$ 0.14 
\\ 
\hline
$\Xi_{bb} \to \Xi^\ast_{bc}$       
& 0.53 & $0.61^{+0.04}$ & 0.25 $\pm$ 0.07 & 
$\Omega_{bb} \to \Omega^\ast_{bc}$ 
& 0.55 & $0.67^{+0.08}$ & 0.29 $\pm$ 0.10 
\\ 
\hline
$\Xi_{bc} \to \Xi^\ast_{cc}$        
& 0.72 & $0.75^{+0.06}$ & 0.64 $\pm$ 0.19 & 
$\Omega_{bc} \to \Omega^\ast_{cc}$  
& 0.74 & $0.76^{+0.13}$ & 0.62 $\pm$ 0.19 \\ 
\hline
$\Xi'_{bc} \to \Xi^\ast_{cc}$        
& 1.70 & $2.33^{+0.16}$ & 2.01 $\pm$ 0.62 & 
$\Omega'_{bc} \to \Omega^\ast_{cc}$  
& 1.83 & $2.36^{+0.33}$ & 1.93 $\pm$ 0.60 \\ 
\hline 
$\Xi^\ast_{bb} \to \Xi_{bc}$        
& 0.28 & $0.35^{+0.03}$ & 0.14 $\pm$ 0.04 & 
$\Omega^\ast_{bb} \to \Omega_{bc}$  
& 0.29 & $0.38^{+0.04}_{-0.02}$ & 0.15 $\pm$ 0.05 \\ 
\hline 
$\Xi^\ast_{bc} \to \Xi_{cc}$        
& 0.38 & $0.43^{+0.06}$ & 0.30 $\pm$ 0.08 & 
$\Omega^\ast_{bc} \to \Omega_{cc}$  
& 0.40 & $0.44^{+0.06}$  & 0.27 $\pm$ 0.07 \\ 
\hline
$\Xi^\ast_{bb} \to \Xi'_{bc}$        
& 0.82 & $1.04^{+0.06}$ & 0.36 $\pm$ 0.10 & 
$\Omega^\ast_{bb} \to \Omega'_{bc}$  
& 0.85 & $1.13^{+0.11}_{-0.08}$ & 0.42 $\pm$ 0.14 \\ 
\hline 
$\Xi^\ast_{bb} \to \Xi^\ast_{bc}$       
& 1.92 & $2.09^{+0.16}$ & 1.05 $\pm$ 0.40 & 
$\Omega^\ast_{bb} \to \Omega^\ast_{bc}$ 
& 2.00 & $2.29^{+0.31}_{-0.04}$ & 1.11 $\pm$ 0.44 \\ 
\hline 
$\Xi^\ast_{bc} \to \Xi^\ast_{cc}$        
& 2.69 & $2.63^{+0.40}$ & 2.66 $\pm$ 0.86 & 
$\Omega^\ast_{bc} \to \Omega^\ast_{cc}$  
& 2.88 & $2.79^{+0.60}$ & 2.51 $\pm$ 0.81 \\ 
\hline 
\end{tabular}
\end{center}
\end{table}

\begin{table}
\begin{center}
{\bf Table 5.} 
Detailed analysis of semileptonic decay widths of double heavy baryons 
in units of $10^{-14}$ GeV. 

\vspace*{.5cm}
\def\arraystretch{1.1}
\begin{tabular}{|c|c|c|c|c|}
\hline 
& \multicolumn{2}{c|}{Exact results} & & \\
\cline{2-3}
   \,\, Decay mode \,\,  
&  \,\, $e, \mu$ modes \,\, 
&  \,\, $\tau$ mode \,\, 
&  \,\, HQS limit \,\, 
&  \,\, NQM \,\, \\[2mm]
\hline
$\Xi_{bb} \to \Xi_{bc}$        
& 0.80 $\pm$ 0.30 & 0.46 $\pm$ 0.13 & 1.33 $\pm$ 0.61 & 9.58 \\
\hline 
$\Omega_{bb} \to \Omega_{bc}$ 
& 0.86 $\pm$ 0.32 & 0.49 $\pm$ 0.14 & 1.92 $\pm$ 1.15 & 9.69 \\
\hline
$\Xi_{bc} \to \Xi_{cc}$        
& 2.10 $\pm$ 0.70 & 0.97 $\pm$ 0.22 & 4.01 $\pm$ 1.21 & 8.57 \\ 
\hline 
$\Omega_{bc} \to \Omega_{cc}$  
& 1.88 $\pm$ 0.62 & 0.80 $\pm$ 0.16 & 4.12 $\pm$ 1.10 & 7.73 \\ 
\hline
$\Xi'_{bc} \to \Xi_{cc}$       
& 1.10 $\pm$ 0.32 & 0.46 $\pm$ 0.10 & 1.94 $\pm$ 0.50 & 4.74 \\
\hline 
$\Omega'_{bc} \to \Omega_{cc}$ 
& 0.98 $\pm$ 0.28 & 0.38 $\pm$ 0.08 & 1.96 $\pm$ 0.46 & 4.60 \\ 
\hline 
$\Xi_{bb} \to \Xi'_{bc}$       
& 0.43 $\pm$ 0.12 & 0.20 $\pm$ 0.04 & 0.76 $\pm$ 0.30 & 5.74 \\ 
\hline
$\Omega_{bb} \to \Omega'_{bc}$ 
& 0.48 $\pm$ 0.14 & 0.22 $\pm$ 0.05 & 0.81 $\pm$ 0.32 & 4.87 \\  
\hline
$\Xi_{bb} \to \Xi^\ast_{bc}$       
& 0.25 $\pm$ 0.07 & 0.12 $\pm$ 0.02 & 0.61 $\pm$ 0.15 & 1.39 \\ 
\hline
$\Omega_{bb} \to \Omega^\ast_{bc}$ 
& 0.29 $\pm$ 0.10 & 0.13 $\pm$ 0.03 & 0.57 $\pm$ 0.23 & 1.32 \\
\hline
$\Xi_{bc} \to \Xi^\ast_{cc}$       
& 0.64 $\pm$ 0.19 & 0.20 $\pm$ 0.03 & 1.39 $\pm$ 0.34 & 2.13 \\
\hline 
$\Omega_{bc} \to \Omega^\ast_{cc}$  
& 0.62 $\pm$ 0.19 & 0.20 $\pm$ 0.04 & 1.78 $\pm$ 0.64 & 2.19 \\ 
\hline
$\Xi'_{bc} \to \Xi^\ast_{cc}$        
& 2.01 $\pm$ 0.62 & 0.65 $\pm$ 0.12 & 4.40 $\pm$ 0.99 & 6.68 \\
\hline 
$\Omega'_{bc} \to \Omega^\ast_{cc}$  
& 1.93 $\pm$ 0.60 & 0.65 $\pm$ 0.13 & 4.61 $\pm$  1.10 & 7.34 \\ 
\hline 
$\Xi^\ast_{bb} \to \Xi_{bc}$        
& 0.14 $\pm$ 0.04 & 0.06 $\pm$ 0.02 & 0.25 $\pm$ 0.10 & 1.12 \\
\hline  
$\Omega^\ast_{bb} \to \Omega_{bc}$  
& 0.15 $\pm$ 0.05 & 0.07 $\pm$ 0.02 & 0.26 $\pm$ 0.10 & 1.12 \\ 
\hline 
$\Xi^\ast_{bc} \to \Xi_{cc}$        
& 0.30 $\pm$ 0.08 & 0.13 $\pm$ 0.02 & 0.58 $\pm$ 0.14 & 0.45 \\
\hline 
$\Omega^\ast_{bc} \to \Omega_{cc}$  
& 0.27 $\pm$ 0.07 & 0.11 $\pm$ 0.02 & 0.59 $\pm$ 0.13 & 0.42 \\ 
\hline
$\Xi^\ast_{bb} \to \Xi'_{bc}$        
& 0.36 $\pm$ 0.10 & 0.17 $\pm$ 0.03 & 0.76 $\pm$ 0.30 & 3.08 \\ 
\hline
$\Omega^\ast_{bb} \to \Omega'_{bc}$  
& 0.42 $\pm$ 0.14 & 0.20 $\pm$ 0.05 & 0.80 $\pm$ 0.31 & 2.93 \\ 
\hline 
$\Xi^\ast_{bb} \to \Xi^\ast_{bc}$       
& 1.05 $\pm$ 0.40 & 0.55 $\pm$ 0.16 & 1.62 $\pm$ 0.73 & 1.12 \\
\hline 
$\Omega^\ast_{bb} \to \Omega^\ast_{bc}$ 
& 1.11 $\pm$ 0.44 & 0.59 $\pm$ 0.18 & 1.72 $\pm$ 0.77 & 6.70 \\ 
\hline 
$\Xi^\ast_{bc} \to \Xi^\ast_{cc}$        
& 2.66 $\pm$ 0.86 & 1.00 $\pm$ 0.21 & 4.63 $\pm$ 1.23 & 5.32 \\ 
\hline 
$\Omega^\ast_{bc} \to \Omega^\ast_{cc}$  
& 2.51 $\pm$ 0.81 & 0.98 $\pm$ 0.20 & 4.95 $\pm$ 1.26 & 5.43\\ 
\hline 
\end{tabular}
\end{center}
\end{table}  

\newpage 

\begin{figure}
\centering{\
\epsfig{figure=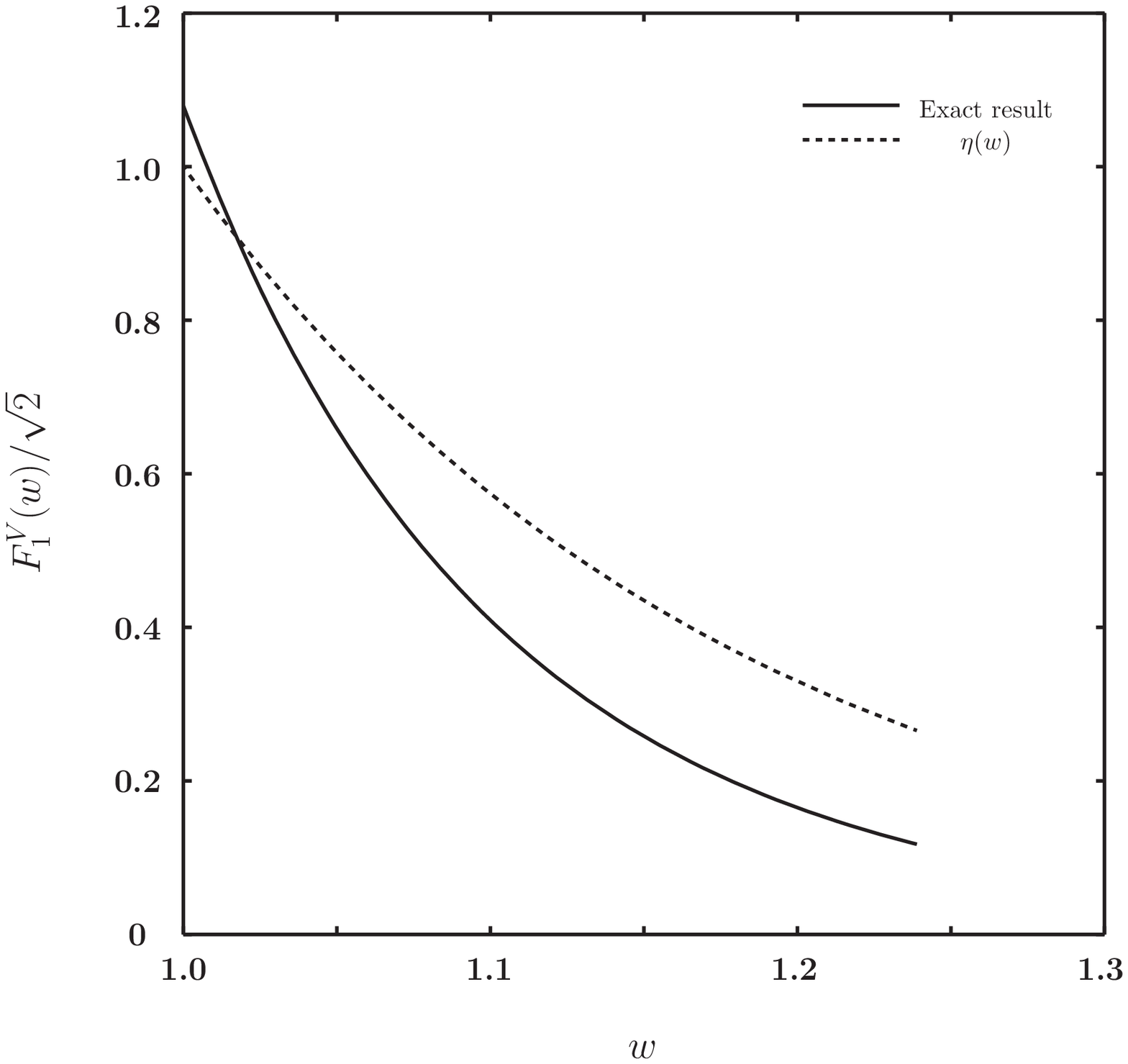,scale=.5}}
\vspace*{9cm}
\caption{Form factor $F_1^V/\sqrt{2}$ describing 
the $\Xi_{bc} \to \Xi_{cc}$ transitions: exact result and 
IW function $\eta(w)$.} 
\label{fig:Fig1}

\vspace*{-1.5cm}

\centering{\
\epsfig{figure=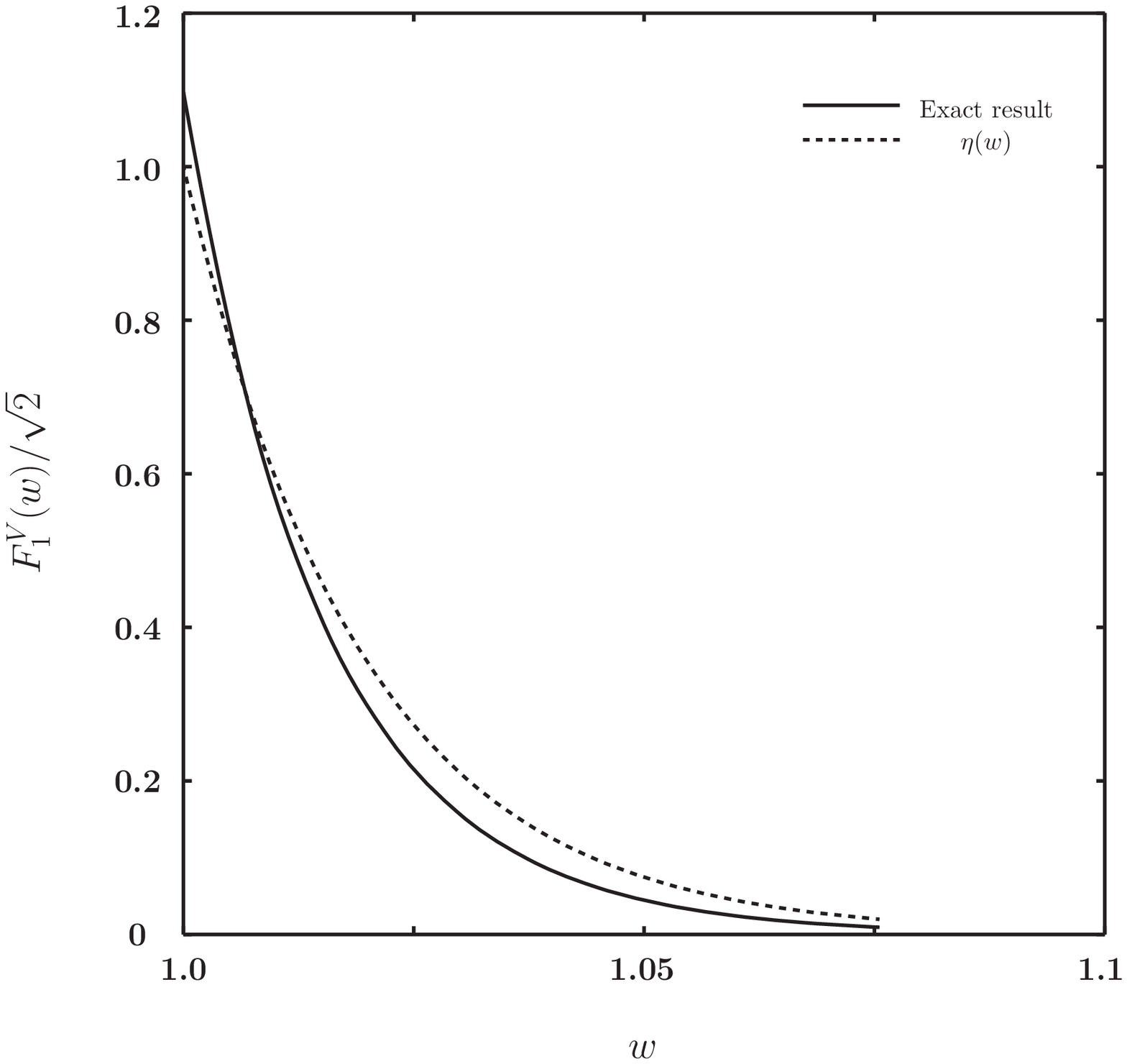,scale=.5}}
\vspace*{9cm}
\caption{Form factor $F_1^V/\sqrt{2}$ describing 
the $\Xi_{bb} \to \Xi_{bc}$ transitions: exact result and 
IW function $\eta(w)$.} 
\label{fig:Fig2}
\end{figure}

\end{document}